\documentclass[journal,compsoc]{IEEEtran}

\usepackage{listings}
\makeatletter
\def\lst@makecaption{%
  \def\@captype{table}%
  \@makecaption
}
\makeatother

\usepackage{todonotes} 
\presetkeys{todonotes}{inline}{}

\usepackage{xspace}
\usepackage{paralist}
\usepackage{graphicx} 
\usepackage{algorithmic}
\usepackage{multirow}
\usepackage{cases}
\usepackage{amsthm}
\usepackage[hidelinks]{hyperref} 
\usepackage{ulem} 
\usepackage{dirtytalk}
\usepackage{balance}
\usepackage{booktabs}

\theoremstyle{definition}
\newtheorem{definition}{Definition}[section]

\ifCLASSOPTIONcompsoc
  \usepackage[nocompress]{cite}
  \usepackage[caption=false,font=normalsize,labelfont=sf,textfont=sf]{subfig}
\else
  \usepackage{cite}
  \usepackage[caption=false,font=footnotesize]{subfig}
\fi

\begin{document}
%
\title{Generating Class-Level Integration Tests Using Call Site Information}

%
%
%
%

\author{Pouria~Derakhshanfar,~\IEEEmembership{Student Member,~IEEE,}
        Xavier~Devroey,~\IEEEmembership{Member,~IEEE,}
        Annibale~Panichella,~
        Andy~Zaidman~\IEEEmembership{Member,~IEEE Computer Society},~
        and~Arie~van~Deursen~\IEEEmembership{Member,~IEEE Computer Society}
}

%
%

\newcommand{\integration}{\textsc{Cling}\xspace}
\newcommand{\cling}{\integration}
\newcommand{\dynaflow}{\textsc{Dyna\-Flow}\xspace}
\newcommand{\evosuite}{\textsc{Evo\-Suite}\xspace}
\newcommand{\randoop}{\textsc{Ran\-doop}\xspace}
\newcommand{\dfj}{\textsc{De\-fe\-cts\-4j}\xspace}
\newcommand{\pit}{\textsc{Pit}\xspace}
\newcommand{\nrun}{20\xspace}

\newcommand\circled[1]{\raisebox{1.2pt}{\textcircled{\hspace{0.35pt}\scriptsize{\raisebox{-.4pt}{#1}}}}}

\newcommand{\eg}{\textit{e.g.,~}}
\newcommand{\Eg}{\textit{E.g.,~}}
\newcommand{\ie}{\textit{i.e.,~}}
\newcommand{\Ie}{\textit{I.e.,~}}
\newcommand{\etal}{\textit{et al.}\xspace}
\newcommand{\etc}{\textit{etc.}\xspace}
\newcommand{\wrt}{\textit{w.r.t.~}}
\newcommand{\cfr}{\textit{cfr.~}}
\newcommand{\Cfr}{\textit{Cfr.~}}
\newcommand{\viz}{\textit{viz.~}}
\newcommand{\aka}{\textit{a.k.a.~}}
\newcommand{\cf}{\textit{cf.~}}
\newcommand{\Cf}{\textit{Cf.~}}
 \markboth{IEEE TRANSACTIONS ON SOFTWARE ENGINEERING,~Vol.~X, No.~X, Date}%
 {Derakhshanfar \MakeLowercase{\textit{et al.}}: Towards Integration-Level Test Case Generation Using Call Site Information}
%



\IEEEtitleabstractindextext{%
\begin{abstract}
  Search-based approaches have been used in the literature to automate the process of creating unit test cases. However, related work has shown that generated tests with high code coverage could be ineffective, i.e., they may not detect all faults or kill all injected mutants. 
  In this paper, we propose \cling, an integration-level test case generation approach that exploits how a pair of classes, the caller and the callee, interact with each other through method calls. In particular, \cling generates integration-level test cases that maximize the Coupled Branches Criterion (CBC).
  Coupled branches are pairs of branches containing a branch of the caller and a branch of the callee such that an integration test that exercises the former also exercises the latter.
  CBC is a novel integration-level coverage criterion, measuring the degree to which a test suite exercises the interactions between a caller and its callee classes. 
  We implemented \cling and evaluated the approach on 140 pairs of classes from five different open-source Java projects. Our results show that (1) \cling generates test suites with high CBC coverage, thanks to the definition of the test suite generation as a many-objectives problem where each couple of branches is an independent objective; 
  (2) such generated suites trigger different class interactions and can kill on average 7.7\% (with a maximum of 50\%) of mutants that are not detected by tests generated randomly or at the unit level; 
  (3) \cling can detect integration faults coming from wrong assumptions about the usage of the callee class (25 for our subject systems) that remain undetected when using automatically generated random and unit-level test suites.
\end{abstract}

}

\maketitle

\IEEEdisplaynontitleabstractindextext

%
\IEEEpeerreviewmaketitle

\IEEEraisesectionheading{\section{Introduction}
\label{sec:introduction}}


\IEEEPARstart{S}{earch-based} approaches have been applied to a variety of white-box testing activities \cite{Harman2012}, among which test case and data generation \cite{McMinn2004}. 
In white-box testing, most of the existing work has focused on the unit level, where the goal is to generate tests that achieve high structural (e.g., branch) coverage.
Prior work has shown that search-based unit test generation can achieve high code coverage~\cite{almasi2017industrial, Campos2017, Panichella2018a}, detect real-bugs~\cite{fraser20151600, Shamshiri2016}, and help developers during debugging activities~\cite{Ceccato2015, Panichella2016}.
 
Despite these undeniable results, researchers have identified various limitations of the generated unit tests~\cite{gay2015risks, Shamshiri2016, schwartz2018}. 
Prior studies have questioned the effectiveness of the generated unit tests with high code coverage in terms of their capability to detect real faults or to kill mutants when using mutation coverage. 
For example, Gay \etal~\cite{gay2015risks} have highlighted how traditional code coverage could be a poor indicator of test effectiveness (in terms of fault detection rate
 and mutation score). Shamshiri \etal~\cite{Shamshiri2016} have reported that around 50\% of faults remain undetected when relying on generated tests with high coverage. 
 Similar results have also been observed for large industrial systems~\cite{almasi2017industrial}. 

Gay \etal~\cite{gay2015risks} have observed that traditional unit-level adequacy criteria only measure whether certain code elements are reached, but not \textit{how} each element is covered. The quality of the test data and the paths from the covered element to the assertion play an essential role in better test effectiveness. As such, they have advocated the need for more reliable adequacy criteria for test case generation tools. While these results hold for generated unit tests, other studies on hand-written unit tests have further highlighted the limitation of unit-level code coverage criteria~\cite{wei2012branch, schwartz2018}.


In this paper, we explore the usage of the integration code between coupled classes as guidance for the test generation process. The idea is that, by exercising the behavior of  a class under test \textit{E} (the calleE) through another class \textit{R} (the calleR) calling its methods, \textit{R} will handle the creation of complex parameter values and exercise valid usages of \textit{E}. In other words, the caller \textit{R} contains integration code that (1) enables the creation of better test data for the callee $E$, and (2) allows to better validate the data returned by \textit{E}.

Integration testing can be approached from many different angles \cite{Jin1998, Offutt2000b}. Among others, \textit{dataflow analysis} seeks to identify possible interactions between the definition and usage (def-use) of a variable. Various coverage criteria based on intra- (for class unit testing) and inter-class (for class integration testing) def-uses have been defined over the years~\cite{Su2017, Harrold1994, Souter2003, Alexander2000, Alexander2010, vivanti2013search, Denaro2015}. Dataflow analysis faces several challenges, including the scalability of the algorithms to identify def-use pairs \cite{Denaro2008} and the number of test objectives that is much higher for dataflow criteria compared to \textit{control flow} ones like branch and branch pair coverage \cite{Su2017, vivanti2013search}. 

In our case, we focus on \textbf{class integration testing} between a caller and a callee~\cite{scott1997building}.  Class integration testing aims to assess whether two or more classes work together properly by thoroughly testing their interactions~\cite{scott1997building}. Our idea is to  complement unit test generation for a class under test by looking at its integration with other classes using \textbf{control flow analysis}. To that end, we define a novel structural adequacy criterion called the \textbf{Coupled Branches Coverage} criterion (CBC), targeting specific integration points between two classes. Coupled branches are pairs of branches $\langle r, e\rangle$, with $r$ a branch of the caller, and $e$ a branch of the callee, such that an integration test that exercises branch $r$ also indirectly exercises branch $e$.

Furthermore, we implement a search-based approach that generates integration-level test suites leveraging the CBC criterion. We name our approach \integration (for \underline{cl}ass \underline{in}tegration testin{\underline g}). \cling uses a state-of-the-art many-objective solver that generates test suites maximizing the number of covered coupled branches. For the guidance, \cling uses novel search heuristics defined for each pair of coupled branches (the search objectives).

We conducted an empirical study on 140 well-distributed (in terms of complexity and coupling) pairs of caller and callee classes extracted from five open-source Java projects. Our results show that \cling can achieve up to 99\% CBC coverage, with an average of 49\% across all pairs of classes. 
We analyzed the benefits of the integration-level test cases generated by \cling compared to unit-level tests generated by \evosuite \cite{Fraser2011}, the state-of-the-art generator of unit-level tests, and random tests generated by \randoop \cite{Pacheco2007}, a random-based test case generator. In particular, we assess whether integration-level tests generated by \cling can kill mutants and detect faults that would remain uncovered when relying on other generated tests given the same generation budget. 

According to our results, on average, \cling kills 7.7\% (resp. 13\%) of the mutants per class that remain undetected by other tests generated using \evosuite (resp. \randoop) for both the caller and the callee. The improvements in mutation score are as high as 50\% for certain pairs of classes. Our analysis indicates that many of the most frequently killed mutants are produced by integration-level mutation operators. 
Finally, we have found 25 integration faults (\ie faults due to wrong assumptions about the usage of the callee class) that were detected only by the integration tests generated with \cling (and not through testing with \evosuite or \randoop).

The remainder of the paper is organized as follows. Section~\ref{sec:background} summarizes the background and related work in the area. 
Section~\ref{sec:approach} defines the Coupled Branches Criteria and introduces \cling, our integration-level test case generator. Section~\ref{sec:evaluation} describes our empirical study, while Section~\ref{sec:results} reports the corresponding empirical results. Section~\ref{sec:discussion} discusses the practical implication of our results. Section~\ref{sec:threats} discusses the threats to validity. Finally, Section~\ref{sec:future-conclusion} concludes the paper.

\section{Background and related work}
\label{sec:background}

McMinn~\cite{McMinn2004} defined search-based software testing (SBST) as \textit{``using a meta-heuristic optimizing search technique, such as a genetic algorithm, to automate or partially automate a testing task"}.
Within this realm, test data generation at different testing levels (such as \textit{unit testing}, \textit{integration testing}, \etc) has been actively investigated~\cite{McMinn2004}. This section 
provides an overview of earlier work in this area.

\subsection{Search-based approaches for unit testing}

SBST algorithms have been extensively used for unit test generation. Previous studies confirmed that such generated tests achieve a high code coverage~\cite{Panichella2018a, Campos2018}, real-bug detection~\cite{almasi2017industrial}, and a debugging cost reduction~\cite{soltani2017, Panichella2016}, complementing manually-written tests.

From McMinn's \cite{McMinn2004} survey about search-based test data generation, we observe that most of the current approaches rely on the control flow graph (CFG) to 
abstract the source code and represent possible execution flows. The $\mathit{CFG_m}=(N_m,E_m)$ represents a method (or function in procedural programming languages) $m$ as a directed graph of \textbf{basic blocks} of code (the nodes $N_m$), while $E_m$ is the set of the control flow edges. An edge connects a basic block $n_1$ to another one $n_2$ if the control may flow from the last statement of $n_1$ to the first statement of $n_2$.

\begin{figure}[t]
    \centering
	\includegraphics[width=0.85\linewidth]{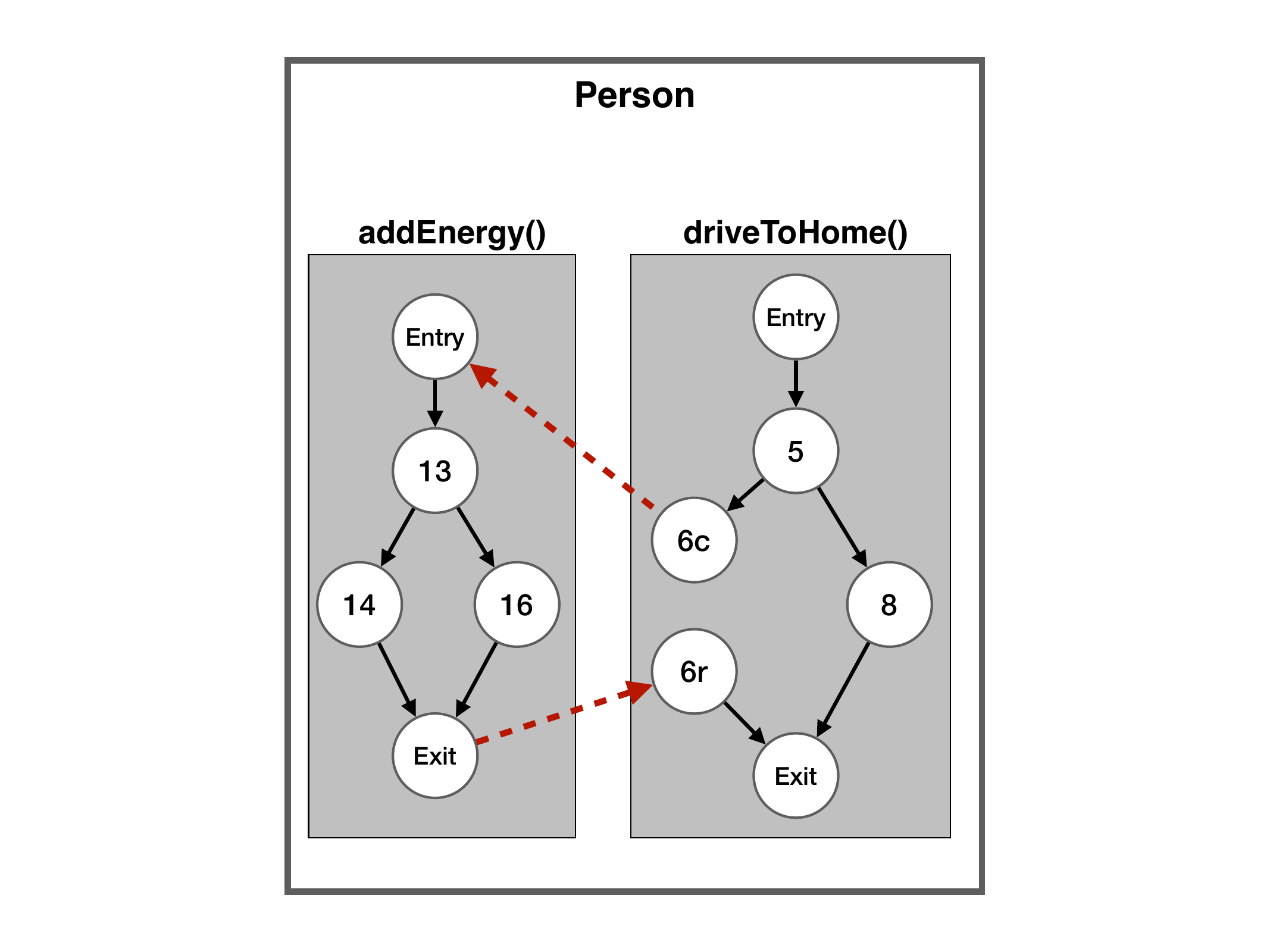}
	\caption{Class-level CFG for class \texttt{Person}}
  \label{fig:CCFG}
\end{figure}

Listing~\ref{list:ClassA} presents the source code of \texttt{Person}, a class representing a person and her transportation habits. A \texttt{Person} can drive home (lines 4-10), or add energy to her car (lines 12-18). Figure~\ref{fig:CCFG} presents the CFG of two of Person's methods, with the labels of the nodes representing the line numbers in the code. Since method \texttt{driveToHome} calls method \texttt{addEnergy}, \textit{node 6} is transformed to two nodes, which are connected to the entry and exit point of the called method. This transformation is explained in the last paragraph of this section. 

\begin{lstlisting}[frame=tb,
    caption={Class \texttt{Person}},
    label=list:ClassA,
    language=java,
    captionpos=t,
    numbers=left,
    belowskip=-2.5em,
    float=t,
    firstnumber=1]
class Person{
    private Car car = new Car();
    protected boolean lazy = false;
    public void driveToHome(){
        if (car.fuelAmount < 100) { (*@\label{line:branchdistance}@*)
            addEnergy();
        } else {
            car.drive();
        }   
    }

    protected void addEnergy(){
        if (this.lazy) {
            takeBus();
        } else {
            car.refuel();
        }
    }   
}
  \end{lstlisting}

Many approaches based on CFGs combine two common heuristics to reach a high branch and statement coverage in unit-level testing: the \textit{branch distance} and the \textit{approach level}.
The \textit{branch distance} measures (based on a set of rules) the distance to \textit{satisfying} (true branch) and the distance to \textit{not satisfying} (false branch) a particular branching node in the program.
For instance, the distance to true for the condition at line \ref{line:branchdistance} in Listing~\ref{list:ClassA} is $100 - \mathit{car.fuelAmount} + 1$, and the distance to false is $\mathit{car.fuelAmount} - 100$.
The \textit{approach level} measures the distance between the execution path and a target node in a CFG. For that, it relies on the concepts of \textbf{post-dominance} and \textbf{control dependency}~\cite{Allen:1970:CFA:800028.808479}.
As an example, in Figure \ref{fig:CCFG}, \textit{node 8} is control dependent on \textit{node 5} and \textit{node 8} post-dominates edge $\langle 5,8\rangle$. The \textit{approach level} is the minimum number of control dependencies between a target node and an executed path by a test case. 


In this study, we analyze how a class is used/invoked by the other classes within the same system. For this purpose, we merge the Class-level Control Flow Graph (CCFG) of target callee and caller classes. 

\subsection{Search-based approaches for integration testing}

Integration testing aims at finding faults that are related to the interaction between components. We discuss existing integration testing criteria and explain the search-based approaches that use these criteria to define fitness functions for automating integration-level testing tasks.

\subsubsection{Integration testing criteria}
\label{sec:background-integ-testing-criteria}

Jin \etal \cite{Jin1998} categorize the connections between two procedures into four types: \textit{call couplings} (type 1) occur when one procedure calls another procedure; \textit{parameter couplings} (type 2) happen when a procedure passes a parameter to another procedure; \textit{shared data couplings} (type 3) occur when two procedures refer to the same data objects; \textit{external device coupling} (type 4) happens when two procedures access the same storage device.
They introduce integration testing criteria according to the data flow graph (containing the definitions and usages of variables at the integration points) of procedure-based software. Their criteria, called \textit{coupling-based testing criteria}, require that the tests' execution paths cover the last definition of a parameter's value in the CFG of a procedure (the \textit{caller procedure}), a node (the \textit{call site}) calling another procedure with that parameter, and the first use of the parameter in the \textit{callee} (and in the caller after the call if the parameter is a call-by-reference). 

Harrold \etal \cite{Harrold1994} introduced data flow testing for a single class focusing on method-integration testing. They define three levels of testing: \textit{intra-method testing}, which tests an individual method (\ie the smallest possible unit to test); \textit{inter-method testing}, in which a public method is tested that (in)directly calls other methods of the same class, and \textit{intra-class testing}, in which the various sequences of public methods in a class are tested. For data flow testing of inter-method and intra-class testing, they defined a \textit{Class-level Control Flow Graph} (CCFG). The CCFG of class \textit{C} is a directed graph $CCFG_C=(N_{Cm},E_{Cm})$ which is a composition of the control flow graphs of methods in $C$; the CFGs are connected through their call sites to methods in the same class~\cite{Harrold1994}. This graph demonstrates all paths that might be crossed within the class by calling its methods or constructors. 

Let us consider again the class $Person$ in Listing~\ref{list:ClassA}. The CCFG of class $Person$ is created by merging the CFGs of its method, as demonstrated in Figure~\ref{fig:CCFF_new}.
For example, in the CFG of the method \texttt{Person.driveToHome()}, the \textit{node 6c} is a call site to \texttt{Person.addEnergy()}. In the approach introduced by Harrold \etal \cite{Harrold1994}, they detect the def-use paths in the constructed CCFGs and try to cover those paths.

Denaro \etal \cite{Denaro2008} revisited previous work on data flow analysis for object-oriented programs \cite{Harrold1994, Souter2003} to define an efficient approach to compute \textit{contextual def-use} coverage \cite{Souter2003} for class integration testing. The approach relies on \textit{contextual data flow analysis} to take state-dependent behavior of classes that aggregate other classes into account. Compared to def-use paths, contextual def-use include the chain of method calls leading to the definition or the use.

A special case is represented by the polymorphic interactions that need to be tested. Alexander \etal \cite{Alexander2000, Alexander2010} used the data flow graph to define testing criteria for integrations between classes in the same hierarchy tree.

All of the mentioned approaches are using data-flow analysis to define integration testing criteria. However, generating data-flow graphs covering the def-uses involved in between classes is expensive and not scalable in complex cases \cite{Su2017}. 
Vivanti \etal \cite{vivanti2013search} shows that the average number of def-use paths in a single class in isolation is three times more than the number of branches. By adding def-use paths between the non-trivial classes, this number grows exponentially. 

In search-based approaches, the number of search objectives matters, as too many objectives lead to search process misguidance. Compared to previous work, our approach does not try to cover def-use paths. Instead, we use a \textit{control flow analysis} to identify from a CCFG a restricted number of pairs of branches (in a caller and a callee) that are not trivially executed together. For instance, the couple of branches $\langle 13,16\rangle$ and $\langle b8,b9\rangle$ in Figure \ref{fig:CCFF_new} are used to define the search objectives of our test case generator. 
Section \ref{sec:approach} details the analysis of the CCFG to identify such pairs of branches, including for special cases of interaction (namely inheritance and polymorphism), and the definition of the objectives and search algorithm.

CCFGs have previously been used in other approaches. For instance, Wang \etal \cite{wang2019could} merge the CFGs of methods of classes in the dependencies of the software under test to identify dependency conflicts.

\subsubsection{Search-based approaches}
\label{subsubsec:search-based-integration-testing}

Search-based approaches are widely used for test ordering \cite{Wang2010, Steindl2012, Hashim2005, Vergilio2012, Bansal2009, JIiang2019, Borner2009, Mariani2016, Guizzo2015, Abdurazik2009, DaVeigaCabral2010, Briand2003a, Vergilio2012}, typically with the aim of executing those tests with the highest likelihood of failing earlier on. 
However, search-based approaches have rarely been used for generating class integration tests. Ali Khan \etal \cite{AliKhan2013} have proposed a high-level evolutionary approach that detects the coupling paths in the data-flow graphs of classes and have used it to define the fitness function for the genetic algorithm. They also proposed another approach for the same goal relying on Particle Swarm Optimization \cite{Khan2014}. 
Since objectives are defined according to the def-use paths between classes, the number of search objectives can grow exponentially, thus severely limiting the scalability of the approach (as we explained in Section \ref{sec:background-integ-testing-criteria}).

Most related to our approach is the  work on \textit{dynamic data flow testing} (\dynaflow) from Denaro \etal \cite{Denaro2015}. Dynamic data flow testing is a two steps \textit{test amplification} pipeline \cite{Danglot2019} where: 
(i) a set of existing test cases are executed to collect execution traces, compute new data flow information, and subsequently derive new test objectives; and 
(ii) the new test objectives are fed to a test case generation tool. The pipeline is repeated until no new test objectives are found. 

In this study, we propose a novel approach for class integration test generation.
Instead of using the data flow graph, which is more expensive to construct than a class call graph, or incrementally amplify the existing test suite, requiring several executions of a test case generation tool, we use the information available in the class call graph of the classes to calculate the fitness of the generated tests. We do note that we could not find any available implementation of data flow-based approaches.

\subsection{Evolutionary approaches for other testing levels}

Arcuri \cite{Arcuri2019} proposed EvoMaster, an evolutionary-based white-box approach for system-level test generation for RESTful APIs. A test for a RESTful web service is a sequence of HTTP requests. EvoMaster tries to cover three types of targets:
 \begin{inparaenum}[(i)]
 \item the statements in the System Under Test (SUT);
 \item the branches in the SUT; and
\item different returned HTTP status codes.
\end{inparaenum}
Although EvoMaster tests different classes in the SUT, it does not systematically target different integration scenarios between classes.

In contrast to EvoMaster, other approaches perform fuzzing \cite{Holler2012}, \textit{``an automated technique providing random data as input to a software system in the hope to expose a vulnerability.''} Fuzzing uses information like grammar specifications \cite{Holler2012, beyene2012, coppit2005, godefroid2008} or feedback from the program during the execution of tests \cite{Padhye2019} to steer the test generation process.
These approaches are black-box and do not rely on any knowledge about classes in the SUT. Hence, their search processes are not guided by the integration of classes.

Our approach performs white-box testing. It monitors the interaction between the target classes and strives to cover different integration scenarios between them.

\section{Class integration testing}
\label{sec:approach}

\begin{figure}[!t]
    \centering
    \includegraphics[width=\columnwidth]{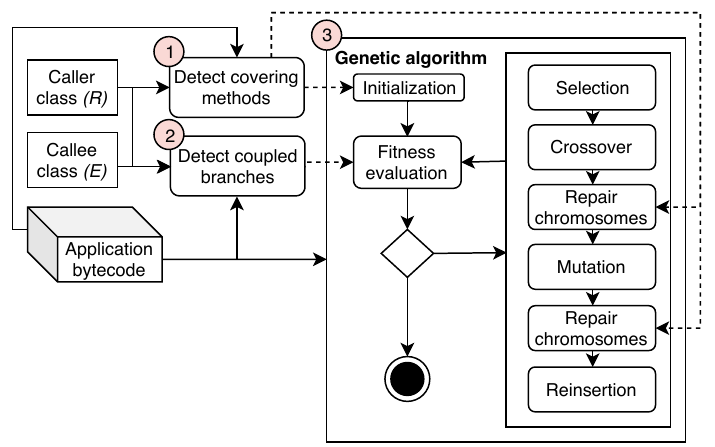}
    \caption{General overview of \integration}
    \label{fig:approach}
\end{figure}

The main idea of our \underline{cl}ass \underline{in}tegration testin\underline{g} (hereinafter referred to as \cling) is to test the integration of two classes by leveraging the usage of one class by another class.
More specifically, we focus on the calls between the former, the callee ($E$), and the latter, the caller ($R$). By doing so, we benefit from the additional context setup by $R$ before calling~$E$ (\eg initializing a complex input parameter), and the additional post-processing after $E$ returns (\eg using the return value later on in $R$), thus (implicitly) making assumptions on the behavior of $E$. 

Figure \ref{fig:approach} presents the general overview of \cling. \cling takes as input a pair of caller-callee $\langle R,E \rangle$ classes with at least one call (denoted \textit{call site} hereafter) from $R$ to $E$. Since the goal of \cling is to generate test cases covering $E$ by calling methods in $R$, the first step (\circled{1}) statically collects the list of \textit{covering methods} in $R$ that, when called, may directly or indirectly cover statements in $E$. This list is later used  during the generation process to ensure that test cases contain calls to covering methods. The second step (\circled{2}) statically analyzes the CCFGs of $R$ and $E$ to identify the coupled branches between $R$ and $E$ used later on to guide the search. 
The CCFGs are statically built from the CFGs of the methods (including inherited ones) in $R$ and $E$.
Finally, the generation of the test cases  (\circled{3}) uses a genetic algorithm with two additional \textit{repair} steps, ensuring that the crossover and mutation only produce test cases able to cover lines in $E$. The result is a test suite for $E$, whose test cases invoke methods in $R$ that cover the interactions between $R$ and $E$.

The remainder of this section describes our novel underlying Coupled Branches Criterion, the corresponding search-heuristics, and test case generation in \cling. 


%

\begin{figure}[t]
	\includegraphics[width=1.0\linewidth]{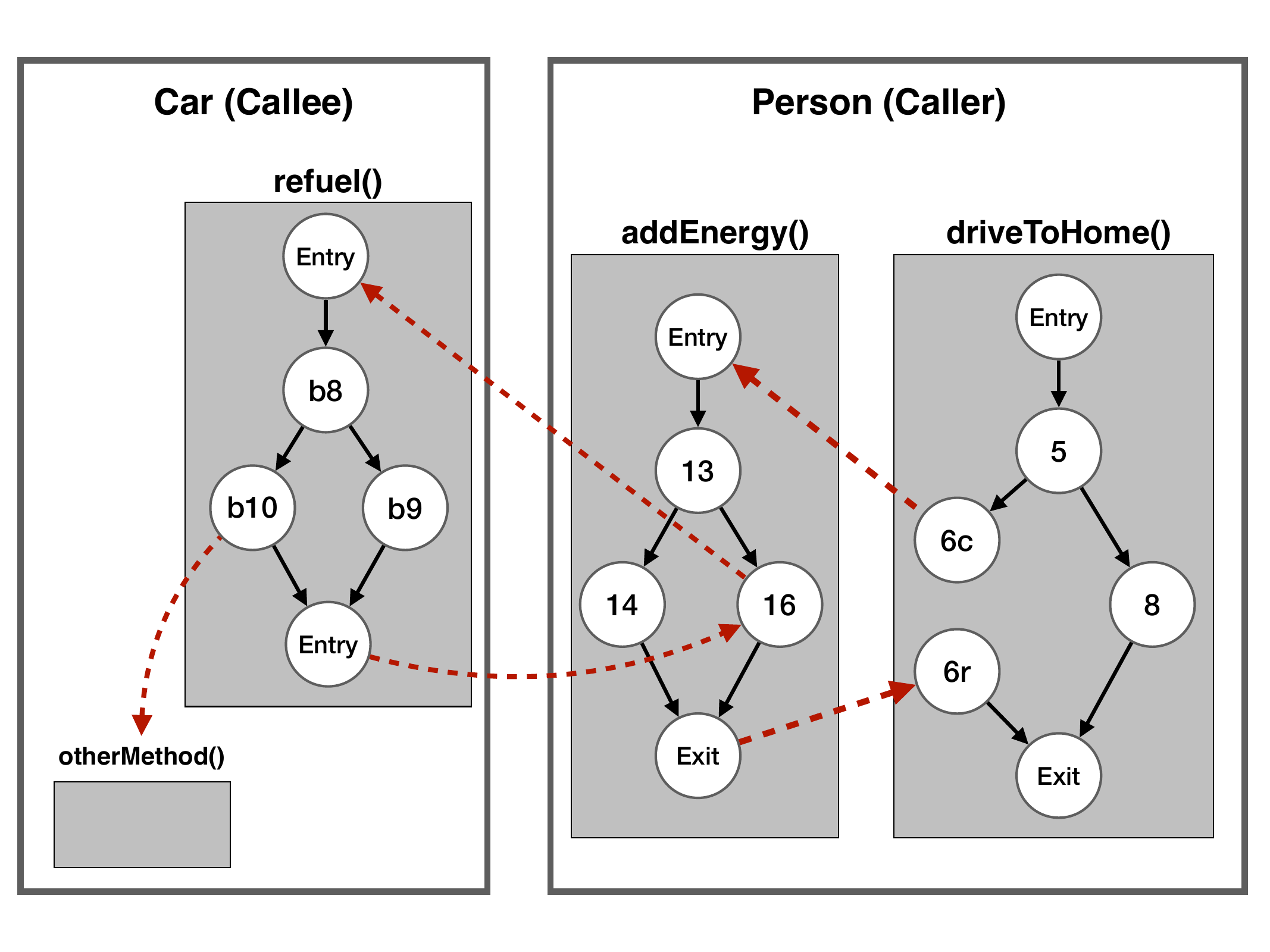}
	\caption{Merging CCFGs of two classes: \texttt{Person} (caller) and \texttt{Car} (callee)}
  \label{fig:CCFF_new}
\end{figure}

\subsection{Coupled Branch testing criterion}

\label{sec:approach:coupledBranchDef}



To test the integration between two classes $E$ and $R$, we need to define a coverage criterion that helps us to measure how thoroughly a test suite $T$ exercises the interaction calls between the two classes ($E$ and $R$). One possible coverage criterion would consist of testing all possible paths (\textit{inter-class path coverage}) that start from the entry node of the caller $R$, execute the integration calls to $E$ and terminate in one of the exit points of $R$. However, such a criterion will be affected by the \textit{path explosion problem}~\cite{Baier2008}: the number of paths increases exponentially with the cyclomatic complexity of $E$ and $R$, and thus the number of interaction calls between the two classes.

To avoid the \textit{path explosion problem}, we define an inte\-gra\-tion-level coverage criterion, namely the Coupled Branch Criterion (CBC), where the number of coverage targets remains polynomial to the cyclomatic complexity of $E$ and $R$. More precisely, CBC focuses on call coupling between caller and callee classes. Intuitively, let $s \in R$ be a call site, \ie a call statement to a method of the class $E$. Our criterion requires to cover all pairs of branches $ (b_r, b_e)$, where $b_r$ is a branch in $R$ that leads to $s$ (the method call), and $b_e$ is a branch of the callee $E$ that is not trivially covered by every execution of $E$. So, in the worst case, the number of coverage targets is quadratic in number of branches in the caller and callee classes.

\subsubsection{Target caller branches} 

Among all branches in the caller class, we are interested in covering the branches that are not trivially (always) executed, and that always lead to the integration call site (\ie calling the callee class) when covered. We refer to these branches as \textit{target branches} for the caller.

\begin{definition}[Target branches for the caller]
    For a call site $s$ in $R$, the set of target branches $B_{R}(s)$ for the caller $R$ contains the branches having the following characteristics:
    \begin{inparaenum}[(i)]
        \item the branches are outgoing edges for the node on which $s$ is control dependent (\ie nodes for which $s$ post-dominates one of its outgoing branches but does not post-dominate the node itself); and
        \item the branches are post-dominated by $s$, \ie branches for which all the paths through the branch to the exit point pass through $s$.
    \end{inparaenum}
\end{definition}

To understand how we determine the target branches in the caller, let us consider the example of the caller and the callee in Figure~\ref{fig:CCFF_new}. The code for the class \texttt{Per\-son} is reported in Listing~\ref{list:ClassA}. The class \texttt{Per\-son} contains two methods, \texttt{add\-Energy()} and \texttt{drive\-To\-Home()}, with the latter invoking the former (line 6 in Listing~\ref{list:ClassA}). The method \texttt{Person.addEnergy()} invokes the method \texttt{refuel()} of the class \texttt{Car} (line 16 in Listing~\ref{list:ClassA}). The method \texttt{Person.driveToHome()} invokes the method \texttt{Car.drive()} (line 8 in Listing~\ref{list:ClassA}). Therefore, the class \textit{Person} is the caller, while \texttt{Car} is the callee. 

Figure~\ref{fig:CCFF_new} shows an excerpt of the Class-level Control Flow Graphs (CCFGs) for the two classes. In the figure, the names of the nodes are labelled with the line number of the corresponding statements in the code of Listing~\ref{list:ClassA}. Node 16 in \texttt{Per\-son.add\-Energy()} is a call site to \texttt{Car.re\-fuel()}; it is also control dependent on nodes 5 (\texttt{Person.drive\-To\-Home()}) and 13 (\texttt{Person.add\-Energy()}). Furthermore, node 16 only post-dominates branch $\langle 13,16\rangle$. Instead, the branch $\langle 5,6c\rangle$ is not post-dominated by  node 16 as covering $\langle 5,6c\rangle$ does not always imply covering node 16 as well. Therefore, the branches in the caller \texttt{Person.addEnergy()} that always lead to the callee are $B_{\mathtt{Person}}(\mathtt{Car.refuel()})=\{\langle 13,16\rangle\}$. 
Hence, among all branches in the caller class (\texttt{Person} in our example), we are interested in covering the branches that, when executed, always lead to the integration call site (i.e., calling the callee class). We refer to these branches as \textit{target branches} for the caller.

\subsubsection{Target callee branches}

Like the target branches of the caller, the target branches of the callee are branches that are not trivially (always) executed each time the method is called. 

\begin{definition}[Target branches for the callee]
    The set of target branches $B_{E}(s)$ for the callee $E$ contains branches satisfying the following properties: \begin{inparaenum}[(i)]
    \item the branches are among the outgoing branches of branching nodes (\ie the nodes having more than one outgoing edge); and 
    \item the branches are accessible from the entry node of the method called in $s$.
    \end{inparaenum}
\end{definition}

Let us consider the example of Figure~\ref{fig:CCFF_new}  again. This time, let us look at the branches in the callee (\texttt{Car}) that are directly related to the integration call. In the example, executing the method call  \texttt{Car.refuel()} (node 16 of the method \texttt{Person.addEnergy()}) leads to the execution of the branching node $b8$ of the class \texttt{Car}. Hence, the set of branches affected by the interaction calls is $B_{\mathtt{Car}}(\mathtt{Car.refu}$-$\mathtt{el()}) = \{\langle b8,b9\rangle;$ $\langle b8,b10\rangle\}$. In the following, we refer to these branches as target branches for the callee. Note that, for a call site $s$ in $R$ calling $E$, the set of target branches for the callee also  includes branches that are trivially executed by any execution of $s$. 


\subsubsection{Coupled branches}
\label{sec:approach:coupledBranches}
Given the sets of target branches for both the caller and  callee, an integration test case should exercise at least one target branch for the caller (branch affecting the integration call) and one target branch for the callee (i.e., the integration call should lead to covering branches in the callee). In the following, we define pairs of target branches $(b_r \in B_{R}(s), b_e \in B_{E}(s))$ as \textit{coupled branches} because covering  $b_r$ can lead to covering $b_e$ as well. 

\begin{definition}[Coupled branches]
    Let $B_{R}(s)$ be the set of target branches in the caller class $R$; let $B_{E}(s)$ be the set of target branches in the callee class $E$; and let $s$ be the call site in $R$ to the methods of $E$. The set of coupled branches $CB_{R,E}(s)$ is the cartesian product of $B_{R}(s)$ and $B_{E}(s)$:
    \begin{equation}\label{def2}
        CB_{R,E}(s) = CB_{R,E}(s) = B_{R}(s) \times B_{E}(s)
    \end{equation}
\end{definition}

In our example of Figure~\ref{fig:CCFF_new}, we have two coupled branches: the branches $(\langle 13,16\rangle,\langle b8,b9\rangle)$ and the branches $(\langle 13,16\rangle$, $\langle b8,b10\rangle)$.

\begin{definition}[Set of coupled branches]\label{def3}
    Let $S=(s_1, \dots, s_k)$ be the list of call sites from a caller $R$ to a callee $E$, the set of coupled branches for $R$ and $E$ is the union of the coupled branches for the different call sites $S$: 
    $$CB_{R,E} = \cup_{s \in S} CB_{R,E}(s)$$
\end{definition}

\subsubsection{The Coupled Branches Coverage criterion (CBC)}\label{sec:approach:coupledBranchCrit}




%
%
Based on the definition above, the CBC criterion requires that for all the call sites $S$ from a caller $R$ to a callee $E$, a given test suite $T$ covers all the coupled branches:
$$
CBC_{R,E} = \frac{
        \vert \{(r_i,e_i) \in CB_{R,E} \vert \exists t \in T: t \ covers \ r_i \ and \ e_i\} \vert
    }{
        \vert CB_{R,E} \vert
    }
$$
%
%
We do note that this formula is only relevant if there are indeed call interactions between caller and callee.
As for classical branch and branch-pair coverage, $CB_{R,E}$ may contain incompatible branch-pairs (\eg when the conditions are mutually exclusive). However, detecting and filtering such pairs is an undecidable problem. Hence, in this study, we target all coupled branches.

\subsubsection{Inheritance and polymorphism}
\label{subsec:inheritance}

\begin{lstlisting}[frame=tb,
    caption={Class GreenPerson},
    label=list:ClassAPrime,
    captionpos=t,
    numbers=left,
    belowskip=-2em,
    float=t,
    language=java,
    firstnumber=1]
class GreenPerson extends Person{
    private HybridCar car = new HybridCar();
    @override
    public void addEnergy(){
        if(this.lazy){
            takeBus();
        }else if (chargerAvailable()){
            car.recharge()
        }else{
            car.refuel();
        }
    }

    private void chargerAvailable(){
        if(ChargingStation.takeavailableStations().size > 0){
            return true;
        }
        return false;
    }
}
\end{lstlisting}


  

In the special case where the caller and callee classes are in the same inheritance tree, we use a different procedure to build the CCFG of the super-class and find the call sites $S$. The CCFG of the super-class is built by merging the CFGs of the methods that are not overridden by the sub-class. As previously, the CCFG of the sub-class is built by merging the CFGs of the methods defined in this class, including the inherited methods overridden by the sub-class (other non-overridden inherited methods are not part of the CCFG of the sub-class).



\begin{figure}[!t]
\vspace{1.5mm}
    \includegraphics[width=0.94\linewidth]{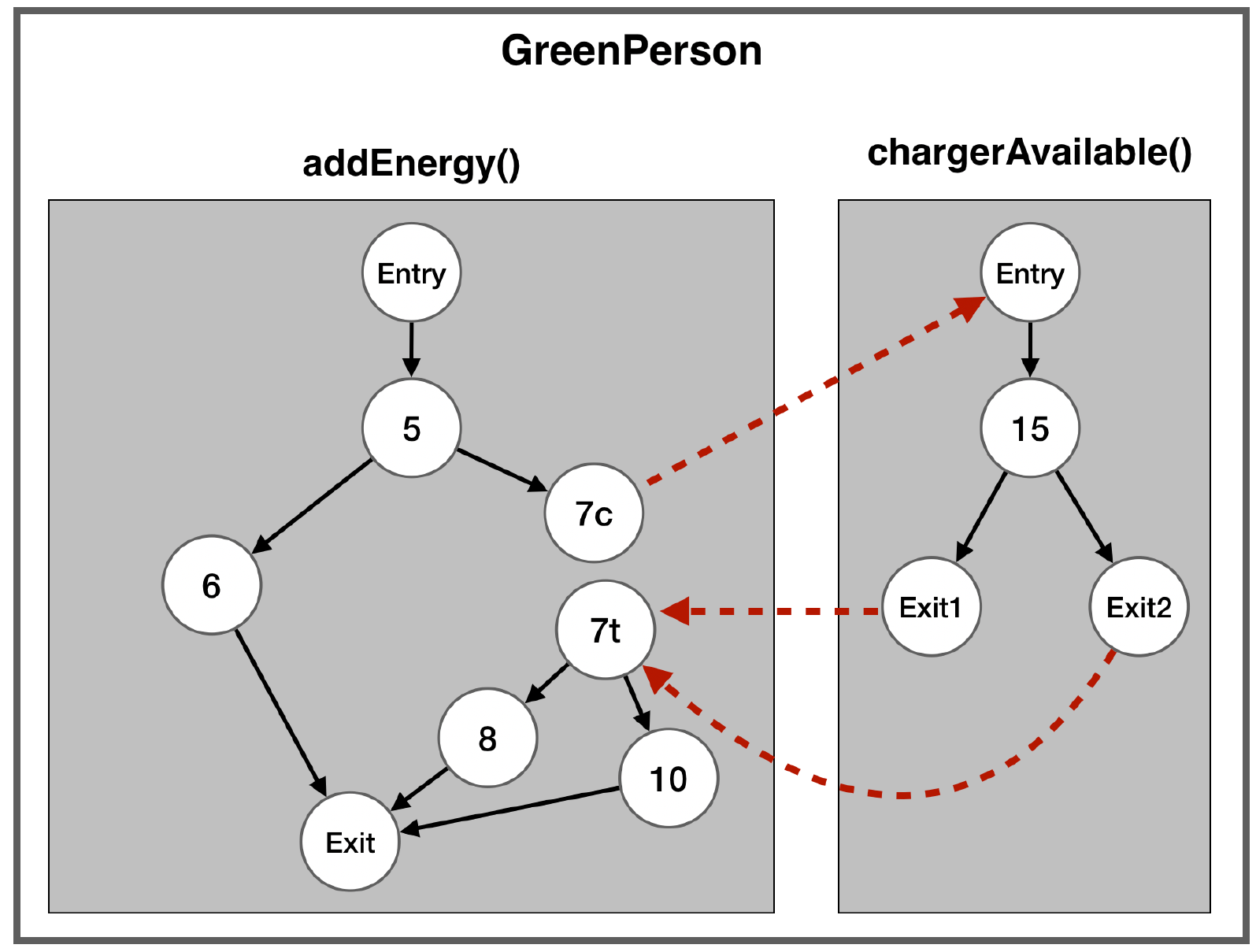}
  \caption{CCFG of $GreenPerson$ as subclass}
  \label{fig:greenPersonCCFG}
\end{figure}

For instance, the class \texttt{GreenPerson} in Listing \ref{list:ClassAPrime}, representing owners of hybrid cars, extends class \texttt{Person} from Listing \ref{list:ClassA}. For adding energy, a green person can either refuel or recharge her car (lines 7 to 11). \texttt{Green\-Person} overrides the method \texttt{Person.add\-En\-er\-gy()} and defines an additional method \texttt{Green\-Person.char\-ger\-Avai\-la\-ble()} indicating whether the charging station is available. Only those two methods are used in the CCFG of the class \texttt{Green\-Person} presented in Figure \ref{fig:greenPersonCCFG}, inherited methods are not included in the CCFG; the CCFG of the super-class \texttt{Person} does not contain the method \texttt{Person.addEnergy()}, redefined by the sub-class \texttt{GreenPerson}.

The call sites $S$ are identified according to the CCFGs, depending on the caller and the callee. If the caller $R$ is the super-class, $S$ will contain all the calls in $R$ to methods that have been redefined by the sub-class. For instance, nodes 6 and 13 in Figure \ref{fig:CCFF_new} with \texttt{Person} as caller. If the caller $R$ is the sub-class, $S$ will contain all the calls in $R$ to methods that have been inherited but not redefined by $R$. For instance, node 7c in Figure \ref{fig:greenPersonCCFG} with \texttt{GreenPerson} as caller.



\subsection{\cling}
\label{sec:cling}

\cling is the tool that we have developed to generate integration-level test suites that maximize the proposed CBC adequacy criterion. The inputs of \cling are the (1) \textit{application's bytecode}, (2) a \textit{caller class} $R$, and (3) a \textit{callee class}~$E$. As presented in Figure \ref{fig:approach}, \cling first detects the covering methods (step \circled{1}) and identifies the coupled branches $CB_{R,E}(s)$ for the different call sites (step \circled{2}), before starting the search-based test case generation process (detailed in the following subsections). \cling produces a test suite that maximizes the CBC criterion for $R$ and $E$.

Satisfying the CBC criterion is essentially a many-objective problem where integration-level test cases have to cover pairs of coupled branches separately. In other words, each pair of coupled branches corresponds to a search objective to optimize. The next subsection describes our search objectives.

\subsubsection{Search objectives}
In our approach, each objective function measures the distance of a generated test from covering one of the coupled branch pairs. The value ranges between $[0,+\infty)$ (zero denoting that the objective is satisfied).
Assume that $CB_{R,E} = \{c_1, c_2, \ldots, c_n \}$ is the set of coupled branches $\langle r_i, e_i \rangle$ between $R$ and $E$. Then, the fitness for a test case $t$ is defined by the following distinct objectives:  

%

\begin{equation}\label{eq:fitness_functions}
    \setlength{\nulldelimiterspace}{0pt}
    Objectives=\left\{
    \begin{IEEEeqnarraybox}[\relax][c]{x}
        $d(c_1,t) = D(r_1,t) \oplus D(e_1,t)$\\
        \dots\\
        $d(c_n,t) = D(r_n,t) \oplus D(e_n,t)$
    \end{IEEEeqnarraybox}\right.
\end{equation}

\noindent where $D(b,t) = al(b,t) + bd(b,t) $ computes the distance between the test $t$ to the branch $b$ using the classical approach level $al(b,t)$ (\ie the minimum number of control dependencies between $b$ and the execution path of $t$) and normalized branch distance $bd(b,t)$ (\ie the distance, computed based on a set of rules, to the branch leading to $b$ in the closest node on the execution path of $t$) \cite{McMinn2004}; and $D(r_i,t)  \oplus D(e_i,t)$ is defined as $D(r_i,t) + 1 $ if $D(r_i,t) > 0$ (\ie the caller branch is not covered) and $D(e_i,t)$ otherwise (\ie the caller branch is covered).

For example, assume that we want to measure the fitness of a test case $t'$, generated during the search process while targeting coupled branches from the classes \texttt{Person} (caller class) and \texttt{Car} (callee class). This test case covers the following path in the CCFG depicted by Figure \ref{fig:CCFF_new}: $Entry \rightarrow 13 \rightarrow 16 \rightarrow b8 \rightarrow b10 \rightarrow Exit$.
As explained in Section \ref{sec:approach:coupledBranches}, this pair of classes contains two coupled branches:  $(\langle 13,16\rangle,\langle b8,b10\rangle)$ and  $(\langle 13,16\rangle,\langle b8,b9\rangle)$, each corresponding to a search objective. Since $t'$ covers both of the branches in the first couple, the objective corresponding to that couple is fulfilled and its fitness value is 0. 
In contrast, $t'$ only covers the first branch of the second couple (\ie $\langle b8,b9\rangle$ is not covered). In this case, $D(r,t')$ equals zero, but $D(e,t')$ is calculated using the approach level and branch distance heuristics. Since $t'$ covers all of the control dependent branches the approach level, $al(b,t')$, equals zero.
The branch distance, $bd(b,t') \in [0,1]$, is calculated according to the concrete values used in the branching condition in the last covered control dependent node (here, $b8$) where the execution path of $t'$ changed away from reaching to the second target branch $\langle b8,b9\rangle$.

\subsubsection{Search algorithm}

To solve such a many-objective problem, we tailored the Many-Objective Sorting Algorithm (MOSA)~\cite{Panichella2015} to generate test cases through class integration. MOSA has been introduced and assessed in the context of unit test generation~\cite{Panichella2015} and security testing~\cite{jan2019search}. Additionally, previous studies~\cite{Campos2018, Panichella2015} have shown that MOSA is very competitive compared with alternative algorithms when handling hundreds and thousands of testing objectives. Interested readers can find more details about the original MOSA algorithm in Panichella \etal~\cite{Panichella2015}. Although a more efficient variant of MOSA has recently been proposed~\cite{Panichella2018}, such a variant (DynaMOSA) requires to have a hierarchy of dependencies between coverage targets that exists only at the unit level. Since targets in unit testing are all available in the same control flow graph, the dependencies between objectives can be calculated (\ie the control dependencies). In contrast, \cling's objective is covering combinations of targets in different control flow graphs. Since covering one combination does not depend on the coverage of another combination, DynaMOSA is not applicable to this problem.

Therefore, in \integration, we tailored MOSA to work at the integration level, targeting pairs of coupled branches rather than unit-level coverage targets (e.g., statements). In the following, we describe the main modifications we applied to MOSA to generate integration-level test cases.
 
%

\subsubsection{Initial population}

The search process starts by generating an initial population of test cases. A random test case is a sequence of statements (\textit{object instantiations}, \textit{primitive statements}, \textit{method calls}, and \textit{constructor calls to the class under test}) of variable lengths. More precisely, the random test cases include \textit{method calls} and \textit{constructors} for the caller $R$, which directly or indirectly invoke methods of the callee $E$ (\textit{covering methods}). Although \integration generates these test cases randomly, it extends the initialization procedure used for search-based crash reproduction \cite{soltani2017}. In particular, the initialization procedure in \cling gives a higher priority to methods in the caller class $R$ that invoke methods of the callee class $E$. While calls to other methods of $R$ are also inserted, their insertion has a lower probability. This prioritization ensures to generate tests covering call sites to the callee class. In the original MOSA algorithm, all methods of the class under test are inserted in each random test case with the same probability without any prioritization. The execution time of the initialisation procedure is part of the search budget.

\subsubsection{Mutation and crossover}\label{sec:mutation_and_crossover} \cling uses the traditional single-point crossover and mutation operators \cite{Fraser2011} (adding, changing and removing  statements) with an additional procedure to repair broken chromosomes. The initial test cases are guaranteed to contain at least one \textit{covering method} (a method of $R$ that directly or indirectly invokes methods of $E$). However, mutation and crossover can lead to generating \textit{offspring} tests that do not include any \textit{covering method}. We refer to these chromosomes as \textit{broken chromosomes}. To fix the broken chromosomes, the \textit{repair procedure} works in two different ways, depending on whether the broken chromosome is created by the crossover or by the mutation. 

If the broken chromosome is the result of the mutation operator, then the repair procedure works as follows: let $t$ be the broken chromosome and let $M$ be the list of covering methods; then, \cling applies the mutation operator to $t$ in an attempt to insert one of the covering methods in $M$. If the insertion is not successful, then the mutation operator is invoked again within a loop. The loop terminates when either a covering method is successfully injected in $t$ or when the number of unsuccessful attempts is greater than a threshold ($50$ by default). In the latter case, $t$ is not inserted in the new population for the next generation.

If the broken chromosome is generated by the crossover operator, then the broken child is replaced by one of its parents.

\subsubsection{Polymorphism}
\label{sec:approach:polymorphism}


If the caller and callee are in the same hierarchy and the caller is the super-class, \cling cannot generate tests for the caller class that will cover the callee class (since the methods to cover are not defined in the super-class). 
This is the case for instance if the super-class (caller) calls abstract methods defined in the sub-class (callee).
In this particular case, \cling generates tests for the callee class. However, it selects the covering methods only from the inherited methods which are not overridden by the callee (sub-class). A covering method should be able to cover calls to the methods that have been redefined by the sub-class. With this slight change, \cling can improve the CBC coverage, as described in Section \ref{subsec:inheritance}.

\subsection{Implementation}
\label{sec:implementation}

We implemented \cling as an open-source tool written in Java.\footnote{Available at \url{https://github.com/STAMP-project/botsing/tree/master/cling}} The tool relies on the \evosuite \cite{Fraser2011} library as an external dependency. It implements the code instrumentation for pairs of classes, builds the CCFGs at the byte-code level, and derives the coverage targets (pairs of branches) according to the CBC criterion introduced in Section~\ref{sec:approach:coupledBranchCrit}. The tool also implements the search heuristics, which are applied to compute the objective scores as described in Section~\ref{sec:approach}. 
 Besides, it implements the repair procedure described in Section~\ref{sec:mutation_and_crossover}, which extends the interface of the genetic operators in \evosuite. 
 Moreover, we customized the many-objective MOSA algorithm~\cite{Panichella2018}, which is implemented in \evosuite, for our test case generation problem in \cling.

\section{Empirical evaluation}
\label{sec:evaluation}

Our evaluation aims to answer three research questions. The first research question analyzes the levels of CBC coverage achieved by \cling. For this research question, we first analyze the coupled branches covered by \cling in each of the cases:
\begin{itemize}
		\item[\textbf{RQ1.1}] \textit{What is the CBC coverage achieved by \cling?}
\end{itemize}
As explained in Section \ref{subsubsec:search-based-integration-testing}, to the best of our knowledge, there is no class-integration test case generator available for comparison. We thus compare \cling to the state-of-the-art unit test generators in terms of CBC coverage:
\begin{itemize}
		\item[\textbf{RQ1.2}] \textit{How does the CBC coverage achieved by \cling compare to automatically generated unit-level and random tests?}
\end{itemize}
Since the test cases generated by \cling aim to cover coupled branches between two classes, we need to determine the effectiveness of this kind of coverage compared to test suites generated for high branch coverage in unit testing:
\begin{itemize}
	\item[\textbf{RQ2.1}] \textit{What is the effectiveness of the integration-level tests compared to unit-level and random tests?}
\end{itemize} 
Additionally, as the integration code of the caller can help to create better test data for the callee and validate its returned data, we investigate the complementarity between \cling and unit testing \wrt fault detection in the callee: 
\begin{itemize}
	\item[\textbf{RQ2.2}] \textit{How complementary are the integration-level tests to the unit-level and random tests \wrt fault detection?}
\end{itemize} 
Finally, we want to see whether the tests generated by \cling can make any difference in practice. Hence, we analyzed the integration faults captured by these tests:
\begin{itemize}
	\item[\textbf{RQ3}] \textit{What integration faults does \cling detect?}
\end{itemize}

\subsection{Baseline Selection}

The goal of this evaluation is to explore the impact and complementarity of the tests generated by \integration on the results of the search-based unit testing in various aspects.
To achieve this purpose, we run our tool against \evosuite, which is currently the best tool in terms of achieving branch coverage \cite{rueda2016unit, panichella2017java, molina2018java, kifetew2019java, Devroey2020}. Additionally, we compare \cling against randomly generated tests using  \randoop~\cite{Pacheco2007}, a feedback-directed random test case generator. In contrast to \evosuite, \randoop can randomly generate tests for multiple classes.

\subsection{Subjects Selection}
\label{sec:setup:subselection}

\begin{table*}[t]
	\center
	\caption{Projects in our empirical study. \texttt{\#} indicates the number of caller-callee pairs. \texttt{CC} indicates the cyclomatic complexity of the caller and callee classes. \texttt{Calls} indicates the number of calls from the caller to the callee. \texttt{Coupled branches} indicates the number of coupled branches.}
	\label{tab:projects}

\begin{tabular}{ l r | rr | rr | rr | rrrr }
\toprule 
\textbf{Project} & \textbf{\#} & \multicolumn{2}{c}{\textbf{Caller}} & \multicolumn{2}{c}{\textbf{Callee}} & \multicolumn{2}{c}{\textbf{Calls}} & \multicolumn{4}{c}{\textbf{Coupled branches}} \\ 
 &  & $\overline{cc}$ & $\sigma$ & $\overline{cc}$ & $\sigma$ & $\overline{count}$ & $\sigma$ & \textit{min} & $\overline{count}$ & $\sigma$ & \textit{max} \\  
 \midrule 
closure & 26 & 1,221.3 & 1,723.0 & 377.2 & 472.5 & 70.3 & 101.0 & 4 & 10,542 & 17,080 & 60,754 \\ 
mockito & 20 & 115.3 & 114.4 & 127.8 & 113.2 & 39.5 & 64.9 & 0 & 1,185 & 1,974 & 6,929 \\ 
time & 51 & 68.7 & 84.0 & 87.2 & 92.3 & 23.9 & 50.5 & 0 & 494 & 1,093 & 5,457 \\ 
lang & 18 & 145.0 & 177.8 & 235.3 & 242.7 & 12.4 & 14.6 & 2 & 409 & 598 & 1,826 \\ 
math & 25 & 79.2 & 88.4 & 57.5 & 64.4 & 18.8 & 34.5 & 2 & 294 & 613 & 2,682 \\ 
\midrule 
\textbf{All} & 140 & 301.1 & 859.5 & 160.6 & 257.7 & 32.4 & 62.8 & 0 & 2,412 & 8,294 & 60,754 \\ 
\bottomrule
\end{tabular}
 
\end{table*}

The subjects of our studies are five Java projects listed in Table \ref{tab:projects}, namely \textit{Closure compiler}, \textit{Apache commons-lang}, \textit{Apache commons-math}, \textit{Mockito}, and \textit{Joda-Time}. 
Our primary reason to use these projects is that they have been used in prior studies to assess the coverage and the effectiveness of unit-level test case generation \cite{ma2015grt, Panichella2018, just2014defects4j, Shamshiri2016}, program repair \cite{martinez_automatic_2017, martinez2016astor}, fault localization \cite{pearson2017evaluating, b2016learning}, and regression testing \cite{noor2015similarity}. A consequence of this selection is that the source code under analysis is relatively old, making it hard to interact with developers to get confirmation about potential faults. Thus, the route that we take instead is to use \textit{future commits} (after the commits under analysis) to explore whether the bugs we identify were addressed (possibly after failures in production), as explained in the next section.

To sample the classes under test, we first extract pairs of \texttt{caller} and \texttt{callee} classes (\ie pairs with interaction calls) in each project. Then, we remove pairs that contain trivial classes, \ie classes where the caller and callee methods have no decision point (\ie with cyclomatic complexity equal to one). This is because methods with no decision points can be covered with single method calls at the unit testing level.
Note that similar filtering based on code complexity has been used and recommended in the related literature~\cite{Campos2017, molina2018java, Panichella2018}. From the remaining pairs, we sampled 140 distinct pairs of classes from the five projects in total, which offers a good balance between generalization (\ie the number of pairs to consider) and statistical power (\ie the number of executions of each tool against each class or pair of classes). 
We performed the sampling to have classes with a broad range of complexity and coupling. In our sampling procedure, each selected class pair includes either the classes with the highest cyclomatic complexity or the mosts coupled classes.
The numbers of pairs selected from each project are reported in Table~\ref{tab:projects}  (column \texttt{\#}), as well as the average cyclomatic complexity ($\overline{cc}$) of the caller and the callee, the average number ($\overline{count}$) of calls from the caller to the callee, and the minimum ($min$), average ($\overline{count}$), and maximum ($max$) number of coupled branches.  
Each pair of caller and callee classes represents a target for \cling.

As reported in Table~\ref{tab:projects}, \cling did not identify any coupled-branches for three pairs of classes (one in \texttt{mockito} and two in \texttt{time}). This is due to the absence of target branches in either the caller or the callee, resulting in no couple of branches to cover. Those three pairs have been excluded from the results.

Our replication package \cite{Derakhshanfar2020} 
contains the list of class pairs sampled for our study, their detailed statistics (\ie cyclomatic complexity and the number of interaction calls), and the project versions.

\subsection{Configurations}

To answer the research questions, we run \cling on each of the selected class pairs. 
For each class pair targeted with \cling, we run \evosuite with the caller and the callee classes as target classes under test (\ie each class is targeted independently) to compare the class integration test suite with unit level test suites for the individual classes. 
We configure \evosuite to use \textit{DynaMOSA} (\texttt{-Dal\-go\-rithm=DynaMOSA}), which has the best outcome in structural and mutation coverage \cite{Panichella2018} and branch coverage (\texttt{-Dcri\-terion=BRAN\-CH}). 

\randoop does not include any dynamic dependency analysis and requires the user to manually specify the list of classes whose methods, constructors, and fields may appear in a test. Following the guidelines provided in the  \randoop manual\footnote{\url{https://randoop.github.io/randoop/manual/}}, we use the Java dependencies analysis utility (\texttt{jdeps}) to identify direct and indirect dependencies of the caller and callee classes.
As the first step, we recursively collected all of the dependencies for caller and callee classes (as suggested by the \randoop manual). However, after running the first round of the experiment with all of the dependencies, we noticed that using all the indirect dependencies resulted in a large number of \randoop executions not terminating due to infinite test case executions (\randoop did not terminate in 128/140 of cases used in this experiment). As mentioned in the \randoop manual, this scenario occurs when one of the tests generated by \randoop traps in an infinite loop and drives the whole test generation process to get stuck in an infinite loop. 
Hence, we followed another suggestion mentioned in the \randoop manual and limited the depth to 2: \ie for each caller and callee, we provided a list of classes to use in the generated tests containing the caller and the callee, their direct dependencies, and the direct dependencies of those dependencies. Additionally, we specify (using option \texttt{--require-covered-classes=}) to keep only test cases in which the caller or callee class are directly or indirectly used. As mentioned by the manual, this option only works if \randoop is executed using the \texttt{covered-class javaagent} to instrument the classes. So, we also used this javaagent for \randoop executions in our experiment.

This results in having the following configurations, each one corresponding to a test suite generated by one independent execution of \cling, \randoop or \evosuite:
\begin{compactenum}
\item $T_{\cling}$, the integration-level test suite generated by \cling (\texttt{-target\_classes <Caller>, <Callee>}); 
\item $T_{Ran}$, the random test suite generated by \randoop for the caller and callee (\texttt{--classlist=<Cal\-ler>, <Cal\-lee>, <le\-vel 1 de\-pen\-den\-cies>, <le\-vel 2 de\-pen\-den\-cies>}); 
\item $T_{EvoR}$, the unit-level test suite generated by \evosuite for the caller (\texttt{-class <Caller>}); 
\item $T_{EvoE}$, the unit-level test suite generated by \evosuite for the callee (\texttt{-class <Callee>}).
\end{compactenum}
All other parameters were left to their default values.

\subsection{Evaluation Procedure}

To address the random nature of the three tools, we repeat each run \nrun times (140 pairs of classes $\times$ 4 executions $\times$ \nrun repetitions $=$ 11,200 executions). Moreover, each \cling run is configured with a search budget of five minutes, including two minutes of search initialization timeout. To allow a fair comparison, we run \evosuite for five minutes on each caller and callee class, and \randoop for ten minutes as it generates tests for both the caller and the callee, including default initialization timeout. This represents a total of $\sim$48.6 days execution time for test case generation.

For \textbf{RQ1}, we analyze the CBC coverage achieved by $T_{Cling}$. As the CBC coverage of $T_{EvoE}$ is equal to 0.0 by construction, we compare $T_{Cling}$ with $T_{Ran}$ and $T_{EvoR}$ across the \nrun independent runs.

For \textbf{RQ2}, we measure the effectiveness of the generated test suite using both \textit{line coverage} and \textit{mutation analysis} on the callee classes $E$ (considered as the class under test in our approach). Mutation analysis is a high-end coverage criterion, and mutants are often used as substitutes for real faults since previous studies highlighted its significant correlation with fault-detection capability~\cite{just2014mutants, andrews2005mutation}. Besides, mutation analysis provides a better measure of the test effectiveness compared to more traditional coverage criteria~\cite{wei2012branch} (\eg branch coverage).

We compute the line coverage and mutation scores achieved by $T_{\cling}$ for the callee class in each target class pair. Then, we compare them to the line coverage and mutation scores achieved by $T_{Ran}$, and the unit-level test suites $T_{EvoR}$ and $T_{EvoE}$) for the callee class. Moreover, we analyse the orthogonality of the sets of mutants in the callee that are strongly killed by $T_{\cling}$, and those killed by the random and unit-level tests individually. In other words, we look at whether $T_{\cling}$ allows killing mutants that are not killed at unit-level or by random tests (strong mutation). Also, we analyze the type of the mutants which are only killed by $T_{\cling}$.

For line coverage and mutation analysis, we use \pit~\cite{Coles2016}, which is a state-of-the-art mutation testing tool for Java code, to mutate the callee classes. \pit also collects and reports the line coverage of the test suite on the original class before mutation. \pit has been used in literature to assess the effectiveness of test case generation tools~\cite{DBLP:journals/stvr/ZhuPZ18,ma2015grt, panichella2017java, molina2018java, kifetew2019java, Devroey2020}, and it has also been applied in industry\footnote{http://pitest.org/sky\_experience/}. In our study, we use \pit v.1.4.9 with all mutation operators activated (\ie the \texttt{ALL} mutators group).

For \textbf{RQ3}, we analyze the exceptions triggered by both integration, random, and unit-level test suites. In particular, we extract unexpected exceptions causing crashes, \ie exceptions that are triggered by the test suites but that are 
(i) not declared in the signature of the caller and callee methods using \texttt{throws} clauses, 
(ii) not caught by a \texttt{try-catch} blocks, and
(iii) not documented in the \texttt{Javadoc} of the caller or callee classes.  Then, we manually analyze unexpected exceptions that are triggered by the integration-level test cases (\ie by \cling), but not by the random and unit-level tests. Since our subjects are selected from \dfj, and thereby the projects used as subjects in this study are not the latest versions, the faults that we find for this research question may be fixed in the subsequent commits. Hence, the three first authors performed a code history analysis by looking at the modifications made to the source code of the classes involved in a fault. Based on this analysis, the authors could examine whether the faults found by \cling were later identified, approved, and fixed by the developers.

The test suites generated by \cling, \evosuite, and \randoop may contain \textbf{flaky tests}, \ie test cases that exhibit intermittent failures if executed with the same configuration. To detect and remove flaky tests, we ran each generated test suite five times. 
Hence, the test suites used to answer our three research questions likely do not contain flaky tests. In this process, we identified 8\%, 3.5\%, and 4.7\% of the tests generated by \cling, \evosuite, and \randoop, respectively, as flaky. For \nrun runs, we detected a total of 1,410,320 flaky tests out of 29,785,260 generated test cases.

To keep the execution time (which includes test generation, flaky test detection, and mutation and coverage analysis) manageable, we used a cluster (with 20 CPU-cores, 384 GB memory, and 482 GB hard drive) to parallelize the execution for our evaluation (50 simultaneous executions). With this parallelization, the automated execution of the whole evaluation took about five days (one day for test generation and four days for flaky test detection and mutation and line coverage measurement).

\section{Evaluation Results}
\label{sec:results}

This section presents the results of the evaluation and answers the research questions.

\subsection{CBC achieved by \cling (RQ1.1)}

\begin{figure}[t]
    \centering
    \includegraphics[width=0.48\textwidth]{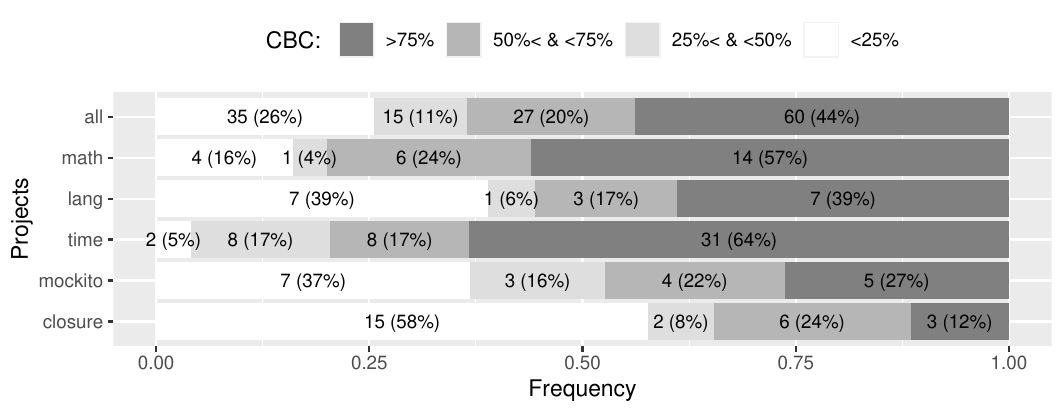}
    \caption{Distribution of \cling's CBC coverage for the different class pairs.}
    \label{fig:cbcdistribution}
\end{figure}

\begin{figure}[t] 
    \centering
    \includegraphics[width=0.48\textwidth]{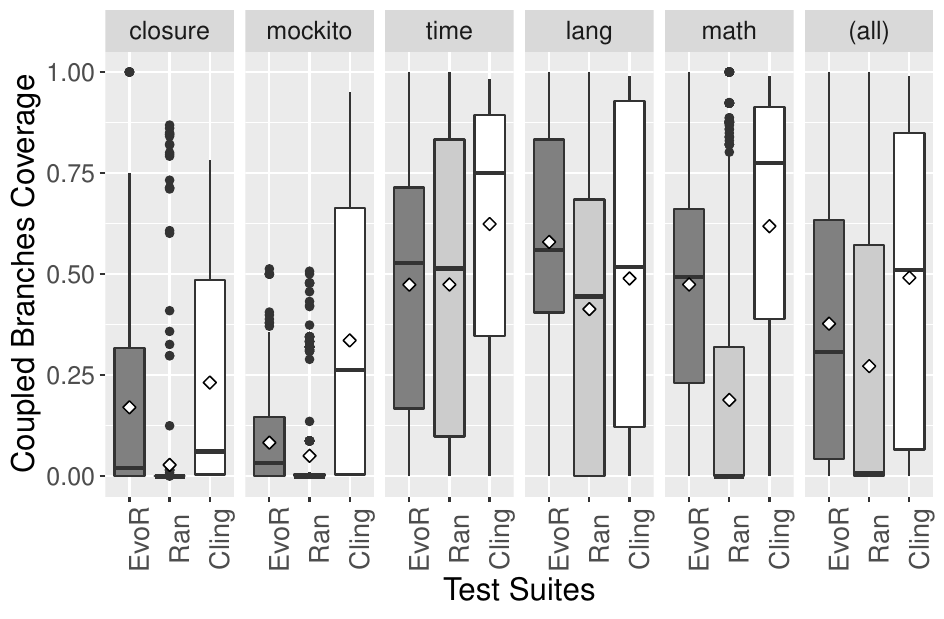}
    \caption{Total coupled-branches coverage achieved by $T_{\cling}$ (Cling), $T_{Ran}$ (Ran) and $T_{EvoR}$ (EvoR). ($\diamond$) denotes the arithmetic mean and (---) is the median.}
    \label{fig:cbccoverage}
\end{figure}
 
As explained in Section \ref{sec:setup:subselection}, \cling did not identify any coupled-branches for three pairs of classes. 
Figure~\ref{fig:cbcdistribution} gives the distribution of the CBC coverage achieved by \cling for 137 pairs of classes. In total, \cling could generate at least one test suite achieving a coupled-branches coverage of at least 50\% for 87 out of 137 class  pairs.  
Figure~\ref{fig:cbccoverage} presents the coupled-branches coverage of $T_{\cling}$ in all projects. On average (the diamonds in Figure~\ref{fig:cbccoverage}) the test suites generated by \cling cover 49.1\% of the coupled-branches.

The most covered couples are in the \texttt{time} project ($62.4\%$ on average), followed by \texttt{math} ($61.9\%$ on average) and \texttt{lang} ($48.9\%$ on average). The least covered couples are in the \texttt{closure} ($23.2\%$ on average)  and \texttt{mockito} projects ($33.6\%$ on average), which are also the projects with the highest number of coupled-branches in Table~\ref{tab:projects} (10,542 coupled-branches on average for all the class pairs in \texttt{closure} and 1,185 coupled-branches on average in \texttt{mockito}).

For 9 caller-callee pairs, \cling could not generate a test suite able to cover at least one coupled branch during \nrun executions: 3 pairs from \texttt{math}, 3 pairs from \texttt{mockito}, 2 pairs from \texttt{closure}, and 1 from \texttt{lang}.
In the class pair from \texttt{lang}, \cling could not cover any coupled branch because the callee class (\texttt{StringUtils}) misleads the search process (we detail the explanation in Section \ref{subsec:rq12}).
The remaining 8 pairs cannot be explained solely by the complexities of the caller (with a cyclomatic complexity ranging from 8 to 5,034 for those classes) and the callee (with a cyclomatic complexity ranging from 1 to 2,186) or the number of call sites (ranging from 1 to 177). This calls for a deeper understanding of the interactions between caller and callee around the call sites. In our future work, we plan to refine the caller-callee pair selection (for which we currently looked at the global complexity of the classes) to investigate the local complexity of the classes around the call sites.

\smallskip\noindent\fbox{\parbox{0.475\textwidth}{
\textbf{Summary (RQ1.1).}  
On average, the generated tests by \cling cover 49.1\% of coupled-branches.
In  87 out of 137 (59.2\%) of the pairs, these test suites achieve a CBC higher than 50\%.
}}

\subsection{CBC achieved by \cling vs. unit tests (RQ1.2)}
\label{subsec:rq12}

\begin{figure}[t]
    \centering
    \includegraphics[width=0.48\textwidth]{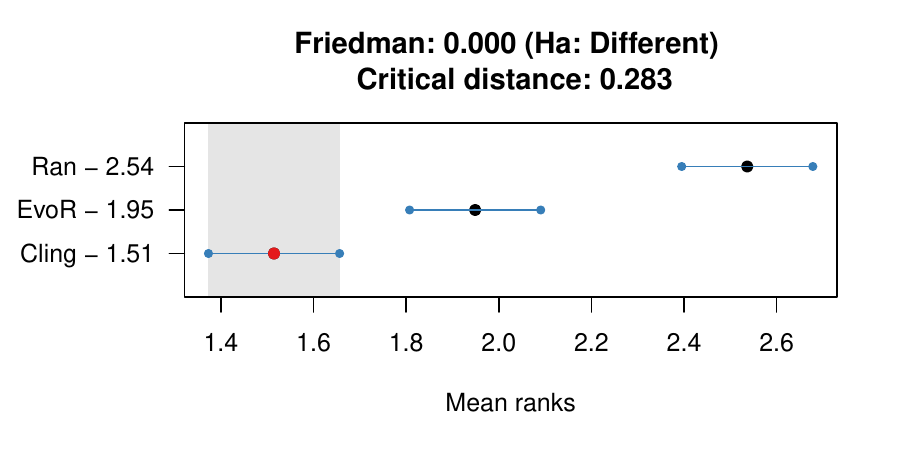}
    \caption{Non-parametric multiple comparisons (\ie mean ranks with confidence interval) in terms of CBC score for $T_{\cling}$ (Cling), $T_{EvoR}$ (EvoR), and $T_{Ran}$ (Ran) using Friedman's test with Nemenyi's post-hoc procedure.}
    \label{fig:cbcfriedman}
\end{figure}

Since $T_{EvoE}$ test suites cover only branches in the callee class (\ie it does not call any methods in the caller class), the coupled-branches coverage achieved by these tests is always zero. 
Hence, for this research question, we compare the tests generated by \cling ($T_{\cling}$) against the tests generated by \randoop ($T_{Ran}$) and \evosuite applied to the caller class ($T_{EvoR}$) \wrt coupled-branches coverage. 

Figure \ref{fig:cbccoverage} presents the coupled-branches coverage of $T_{\cling}$, $T_{Ran}$, and $T_{EvoR}$ for all projects. The number of covered coupled-branches by $T_{\cling}$ is higher in total (\textit{all} in Figure~\ref{fig:cbccoverage}). On average (the diamonds in Figure~\ref{fig:cbccoverage}), the test suites generated by \cling (49.1\%) cover more coupled-branches compared to 37.8\% for $T_{EvoR}$, and 27.2\% for $T_{Ran}$.
%
On average, the coupled-branches coverage achieved by unit tests is lower than the one achieved by \cling in all of the projects except \texttt{lang}. The average coupled-branches coverage of \evosuite in this project is $58\%$, compared to $48.9\%$ for \cling.
We also observe a wider distribution of the CBC coverage for $T_{\cling}$ (with a median of $51.0\%$ and an IQR of $78.2\%$) compared to $T_{EvoR}$ (with a median of $30.7\%$ and an IQR of $59.0\%$) and $T_{Ran}$ (with a median $<1.0\%$ and an IQR of $57.1\%$). 

We further compare the different test suites using Friedman's non-parametric test for repeated measurements with a significance level $\alpha = 0.05$~\cite{Garcia:2009} . This test is used to test  the significance of the  differences between groups (treatments) over the dependent variable (CBC coverage in our case). We complement the test for significance with Nemenyi's post-hoc procedure~\cite{japkowicz2011evaluating,panichella2021systematic}. Figure \ref{fig:cbcfriedman} provides a graphical representation of the ranking (\ie mean ranks with confidence interval) of the different test suites. According to the Friedman test, the different treatments (i.e., \cling, \evosuite, and \randoop) achieve significantly different CBC coverage (p-values $ <0.001$).  According to Figure~\ref{fig:cbcfriedman}, the average rank of \cling is much smaller than the average ranks of the two baselines. Furthermore, the differences between the average rank of $T_{\cling}$ and the average rank of the two baselines are larger than the critical distance $CD=0.283$ determined by Nemenyi's post-hoc procedure. This indicates that $T_{\cling}$ achieves a significantly higher CBC coverage  than $T_{EvoR}$ and $T_{Ran}$.

Finally, we have manually analyzed the search progress of \cling  for pairs of classes where the number of covered coupled-branches is low (\ie lower than 10). 
We noticed that \cling is counter-productive for specific class pairs where the callee class is \texttt{StringUtils}. In those cases, the test cases generated during the search initialization throw a \texttt{NoSuchFieldError} in the callee class (\texttt{StringUtils} here). Since these test cases achieve small approach levels and branch distances from the callee branches, they are fitter (\ie their fitness value is lower) than other test cases. Therefore, these test cases are selected for the next generation and drive the search process in local optima.

\smallskip\noindent\fbox{\parbox{0.475\textwidth}{
\textbf{Summary (RQ1.2).}  
On average, the generated test suites by \cling cover 11.3\% more coupled-branches compared to \evosuite and 21.9\% more coupled-branches compared to \randoop. 
}}

\subsection{Line Coverage and Mutation Scores (RQ2.1)}

\begin{figure*}[t]
    \centering
    \subfloat[Line coverage of the callee (E)]{
        \includegraphics[width=0.48\textwidth]{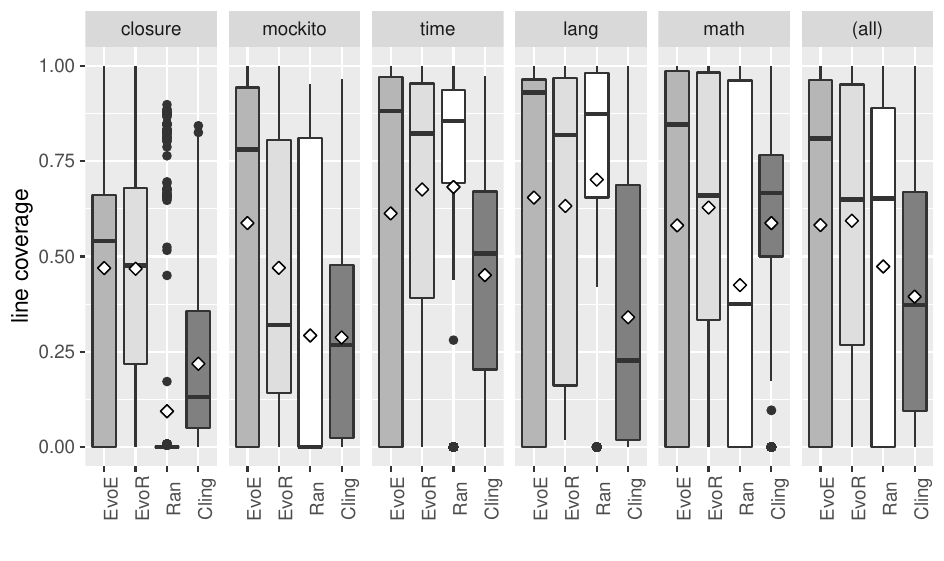}
        \label{fig:linecoverage}
    }
    \hfil
    \subfloat[Mutation score of the callee (E)]{
        \includegraphics[width=0.48\textwidth]{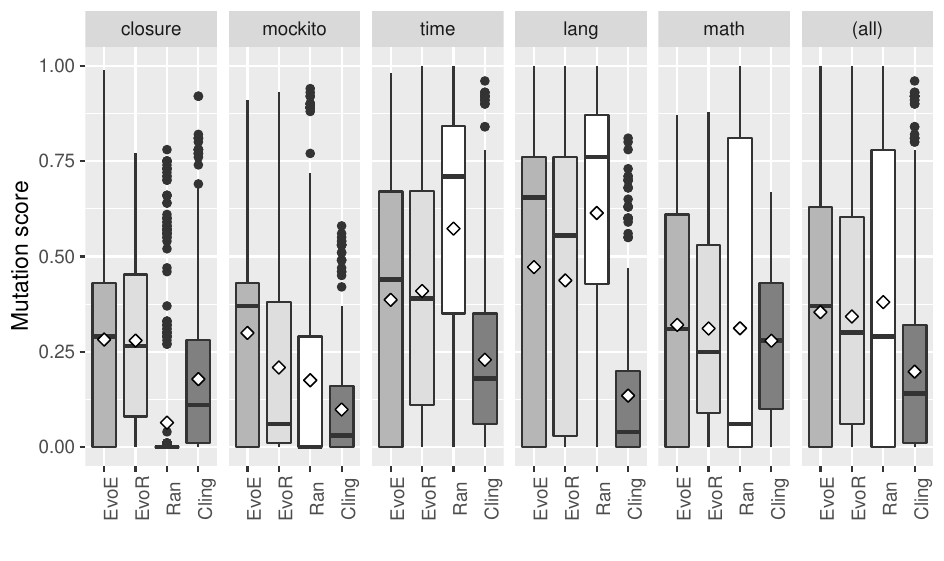}
        \label{fig:mutation:boxplot}
    }
    \caption{Effectiveness of $T_\integration$ (Cling), $T_{EvoE}$ (EvoE), $T_{EvoR}$ (EvoR), and $T_{Ran}$ (Ran). ($\diamond$) denotes the arithmetic mean and (---) indicates the median.}
    \label{fig:effectiveness}
\end{figure*}

Figure \ref{fig:linecoverage} shows line coverage of the callee classes ($E$) for the test suites generated by the different approaches. On average, \cling covers 39.5\% of the lines of the callee classes. This is lower compared to unit-level tests generated using \evosuite (58.2\% for $T_{EvoE}$ and 59.4\% for $T_{EvoR}$), and \randoop (47.4\% for $T_{Ran}$). 

To understand the fault revealing capabilities of \integration compared to unit-level and random test suites, we first show in Figure~\ref{fig:mutation:boxplot} the overall mutation scores when mutating class $E$, and apply the test suite $T_{EvoE}$, $T_{EvoR}$, $T_{Ran}$,  and $T_{\integration}$.
Similar to line coverage, test suites optimized for overall branch coverage achieve a total higher mutation score (35.4\% for $T_{EvoE}$ and 34.2\% for $T_{EvoR}$ on average), simply because a mutant that is on a line that is never executed cannot be killed. \randoop achieves on average the best mutation score (38\% for $T_{Ran}$), which would tend to indicate that despite a lower line coverage, indirect testing of the callee class through its dependencies enables discovering more faults. 
$T_{\integration}$ scores lower (20.0\% on average), since \integration searches for dedicated interaction pairs, but does not try to optimize overall line coverage.
Note that $T_{\integration}$ achieves the highest average mutation score for classes in \texttt{math}, while it achieves the lowest mutation score for classes in the \texttt{mockito} project. 

Our results are consistent with the design and objectives of the tools: \evosuite seeks to cover all the branches of the class under test; \cling targets specific pairs of branches between the caller and callee; and \randoop performs (feedback-directed) random testing. 

\smallskip\noindent\fbox{\parbox{0.475\textwidth}{
\textbf{Summary (RQ2.1).}  
The results in terms of line coverage are as expected, namely that \evosuite has the highest average line coverage (58.2\% for $T_{EvoE}$ and 59.4\% for $T_{EvoR}$), followed by \randoop (47.4\%) and \cling (39.5\%).
Regarding mutation score, \randoop achieved the highest mutation score on average (38\%), followed by \evosuite (35.4\% for $T_{EvoE}$ and 34.2\% for $T_{EvoR}$ on average) and \cling (20.0\%). This tends to indicate that despite a lower line coverage, indirect testing of the callee class through its dependencies in \randoop enables discovering more faults. 
}}

\subsection{Combined Mutation Analysis (RQ2.2)}

\begin{figure}[t]
    \centering
    \includegraphics[width=0.48\textwidth]{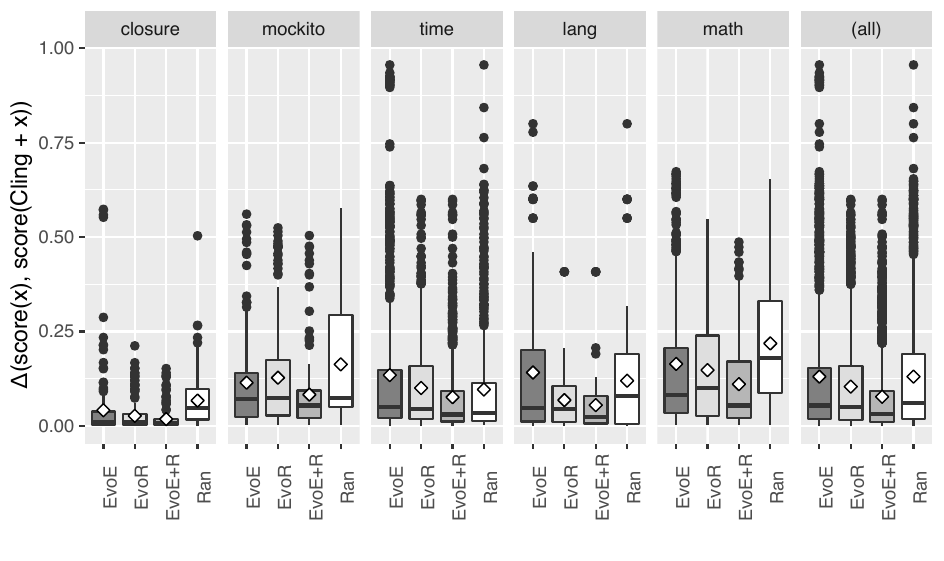}
    \caption{Increases ($\Delta$) of the mutation score when combining $T_\integration$ with unit test suites $T_{EvoE}$ (EvoE),  $T_{EvoR}$ (EvoR), and their unions $T_{EvoE+EvoR}$ (EvoE+R), and $T_{Ran}$ (Ran). ($\diamond$) denotes the arithmetic mean and (---) is the median.}
    \label{fig:mutation:diff:boxplot} 
\end{figure}

Figure~\ref{fig:mutation:boxplot} shows that unit test suites do not kill almost half of the mutants. \integration targets more mutants, including those that remain alive with unit tests.
In Figure \ref{fig:mutation:diff:boxplot}, we report the \textit{improvement} ($\Delta$) in the mutation score when executing $T_\cling$ in addition to different test suites ($T_{EvoE}$, $T_{EvoR}$, and $T_{Ran}$), and their unions ($T_{EvoE+EvoR}$).

On average, 13\% of the mutants are killed only by $T_{\integration}$ compared to both $T_{EvoE}$, the unit test suites optimized for the class under test ($E$), and $T_{Ran}$, randomly generated tests.
This difference decreases to 10.4\% if we use $T_{EvoR}$ the unit test suites exercising $E$ via the caller class $R$ (as more class interactions are executed).  
The difference with traditional unit testing is still 7.7\%, when comparing \integration with the combined unit test suites $T_{EvoE+EvoR}$, exercising $E$ directly as much as possible as well as indirectly via call sites in $R$.

The outliers in Figure~\ref{fig:mutation:diff:boxplot} are also of interest: for 20 classes (out of 137), \cling was able to generate a test suite where more than half of the mutants were killed \emph{only} by $T_{\cling}$, compared to $T_{EvoE}$ (i.e., +50\% of mutation score). When compared to $T_{EvoE+EvoR}$ there are 4 classes for which $T_{\cling}$ kills more than half of the mutants that are killed by neither $T_{EvoE}$ nor $T_{EvoR}$. This further emphasizes the complementarity between the unit and integration testing. Comparing to randomly generated tests $T_{Ran}$, $T_{\cling}$ kills more than half of the mutants for 13 classes, demonstrating the need for guidance when generating class integration tests.  

\begin{table*}[t]
	\center
    \caption{Status (for $T_{EvoR}$, $T_{EvoE}$, and $T_{Ran}$) of the mutants killed solely by $T_{\cling}$. \texttt{Not-covered} denotes the number of mutants killed by $T_{\cling}$, which are not covered by \evosuite (or \randoop) test suites, and \texttt{survived} denotes the number of mutants killed by $T_{\cling}$, which are covered by \evosuite (or \randoop) tests but not killed. The numbers between parentheses denote the percentage of mutants.} 
    \label{tab:mutant-status-table}
    \resizebox{0.99\textwidth}{!}{%
    \begin{tabular}{ l | cc | cc | cc | cc | cc }
\toprule
\textbf{Test Suite}& \multicolumn{2}{c}{\textbf{closure}}& \multicolumn{2}{c}{\textbf{lang}}& \multicolumn{2}{c}{\textbf{math}}& \multicolumn{2}{c}{\textbf{mockito}}& \multicolumn{2}{c}{\textbf{time}} \\ 
 & not-covered & survived & not-covered & survived & not-covered & survived & not-covered & survived & not-covered & survived \\ 
 \midrule 
$T_{EvoE}$&  1,988 ( $<$1\% )&  881 ( $<$1\% )&  3,247 ( 1\% )&  403 ( $<$1\% )&  6,178 ( 5\% )&  1,747 ( 1\% )&  5,604 ( 4\% )&  2,414 ( 2\% )&  10,905 ( 3\% )&  5,920 ( 1\% ) \\ 
$T_{EvoR}$&  2,480 ( $<$1\% )&  780 ( $<$1\% )&  2,797 ( $<$1\% )&  851 ( $<$1\% )&  5,310 ( 4\%   )&  2,558 ( 2\%   )&  4,867 ( 4\%   )&  3,144 ( 2\%   )&  7,431 ( 2\%   )&  9,150 ( 2\%   ) \\ 
$T_{Ran}$&  18,935 ( 2\% )&  246 ( $<$1\% )&  1,834 ( 1\% )&  343 ( $<$1\% )&  13,316 ( 1\% )&  1,381 ( 1\% )&  18,419 ( 1\% )&  949 ( 1\% )&  18,073 ( 4\% )&  5,942 ( 1\% ) \\ 
\bottomrule
\end{tabular}
    }
\end{table*} 

Table \ref{tab:mutant-status-table} presents the status of the mutants that are killed by $T_{\cling}$ but not by unit-level or random test cases. 
What stands out is that many mutants are in fact covered, but not killed by unit-level or random test suites.
Here \cling leverages the context of the caller, not only to reach a mutant, but also to \emph{propagate} the (modified) values inside the caller's context, so that the mutants can be eventually killed.

\subsubsection{Mutation Operators}

\begin{table}[t]
	\center
	\caption{Number of mutants killed solely by $T_{\cling}$ and grouped by mutation operators. Integration-level operators are highlighted in \textbf{bold} face and the corresponding integration-level mutation operator defined by Delamaro \etal \cite{delamaro2001interface} is indicated between parenthesis.}
    \label{tab:mutant-operators-table}
    \resizebox{0.49\textwidth}{!}{%
        \resizebox{0.77\textwidth}{!}{%
\begin{tabular}{l | c r | c r}
    \toprule
    \multicolumn{1}{r}{\textbf{Against}} & \multicolumn{2}{c}{\textbf{EvoSuite}} & \multicolumn{2}{c}{\textbf{Randoop}} \\
    \midrule
    \textbf{Mutation operator} & \textbf{Rank} & \textbf{\#kills} & \textbf{Rank} & \textbf{\#kills} \\ 
    \midrule
    \textbf{NonVoidMethodCallMutator} (\textit{RetStaRep}) & 1 & 1,983 & 1 & 2,340\\ 
    NegateConditionalsMutator & 2 & 1,638 & 2 & 2,020\\ 
    InlineConstantMutator & 3 & 1,201 & 5 & 1,183\\ 
    \textbf{ReturnValsMutator} (\textit{RetStaRep}) & 4 & 1,195 & 4 & 1,414 \\ 
    RemoveConditionalMutator\_EQUAL\_IF & 5 & 1,110 & 3 & 1,424\\ 
    RemoveConditionalMutator\_EQUAL\_ELSE & 6 & 1,015 & 6 & 1,138 \\ 
    \textbf{NullReturnValsMutator} (\textit{RetStaRep}) & 7 & 578 & 7 & 695\\ 
    \textbf{ArgumentPropagationMutator} (\textit{FunCalDel}) & 8 & 518 & 9 & 539\\ 
    MathMutator& 9 & 513 & 11 & 398\\ 
    MemberVariableMutator & 10  & 458 & 8 & 576\\ 
    \textbf{ConstructorCallMutator} (\textit{FunCalDel}) & 11 & 379 & 10 & 481\\ 
    RemoveConditionalMutator\_ORDER\_IF & 12 & 375 & 13 & 348\\ 
    \textbf{VoidMethodCallMutator} (\textit{FunCalDel}) & 13 & 374 & 12 & 394\\ 
    RemoveConditionalMutator\_ORDER\_ELSE & 14 & 348 & 16 & 270\\ 
    ConditionalsBoundaryMutator & 15 & 322 & 15 & 272\\ 
    \textbf{PrimitiveReturnsMutator} (\textit{RetStaRep}) & 16 & 309 & 14 & 295\\ 
    NakedReceiverMutator & 17  & 264 & 17 & 235\\ 
    IncrementsMutator & 18 & 143 & 19 & 154\\ 
    \textbf{BooleanTrueReturnValsMutator} (\textit{RetStaRep}) & 19 & 142 & 18 & 162\\ 
    RemoveIncrementsMutator & 20 & 106 & 22 & 89 \\ 
    RemoveSwitchMutator & 21 & 89 & 20 & 134 \\
    \textbf{EmptyObjectReturnValsMutator} (\textit{RetStaRep}) & 22 & 71 & 21 & 105\\ 
    \textbf{BooleanFalseReturnValsMutator} (\textit{RetStaRep}) & 23  & 63 & 23 & 83 \\ 
    InvertNegsMutator & 24 & 38 & 24 & 44\\ 
    SwitchMutator & 25 & 16 & 25 & 36\\ 
    \bottomrule
    \end{tabular}
}
    
    }
\end{table} 

We analyzed the mutation operators that generate mutants that are exclusively killed by $T_{\cling}$. We categorize the mutation operators implemented in \pit into \textit{integration-level} and \textit{non-integration-level}. For this categorization, we rely on the definition of mutation operators for integration testing provided by Delamaro \etal \cite{delamaro2001interface}. 
We observed that ten of the mutation operators implemented in \pit inject integration-level faults. These operators can be mapped to two integration-level operators defined by Delamaro \etal \cite{delamaro2001interface}: \textit{RetStaRep}, which replaces the return value of the called method, and \textit{FunCalDel}, which removes the calls to void method calls and replaces the non-void method calls by a proper value.

Table \ref{tab:mutant-operators-table} lists the number of mutants killed exclusively by $T_{\cling}$ and grouped by mutation operators. Integration-level operators are indicated in bold with the mapping to either \textit{RetStaRep} or \textit{FunCalDel} between parenthesis. As we can see in this table, the most frequently killed mutants are produced by an integration-level operator, and other integration-level operators also produce frequently killed mutants. 
We can see that all of the ten integration-level mutation operators generate mutants that can be killed using \integration. 

Furthermore, some of the most frequently killed mutants are not produced by integration-level operators. For instance, operator \textit{NegateConditionalsMutator}, which mutates the conditions in the target class, produces the second most frequently killed mutants. These mutants are not killed but also not covered by tests generated by \evosuite. 

\newsavebox\callerbox
\begin{lrbox}{\callerbox}
\begin{minipage}{0.44\textwidth}
    \begin{lstlisting}[
        numbers=left,
        firstnumber=1]
boolean evaluateStepC(StepInterpolator interpolator){
  if (functions.isEmpty()){[...]}
  if (! initialized) {[...]}
  for ([...]) {
      [...];
      // calling the callee class in the next line.
      if (state.evaluateStep(interpolator)){ 
          // Changing variable first
          [...]
      }  
  }
  return first != null;
}
    \end{lstlisting}
\end{minipage}
\end{lrbox}

\newsavebox\calleebox
\begin{lrbox}{\calleebox}
\begin{minipage}{0.44\textwidth}
    \begin{lstlisting}[
        numbers=left,
        firstnumber=1] 
boolean evaluateStep(final StepInterpolator interpolator){
   [...]    
   for([...]){
       if([...]){
           [...];
       }
       if([...]){
           [...];
       }
   }
   [...];
   |return false;| return true; //mutant
}
     \end{lstlisting}
\end{minipage}
\end{lrbox}

\begin{figure*}[t]
    \centering
    \subfloat[Method \texttt{evaluateStepC} declared in the caller class \texttt{SwitchingFunctionsHandler}.]{\usebox\callerbox\label{fig:mutant:example:caller}}
    \hfil
    \subfloat[Mutant \texttt{evaluateStep} declared in the callee class \texttt{switchsState}.]{\usebox\calleebox\label{fig:mutant:example:callee}}
    \caption{Example of a integration-level mutant killed only by \integration From \texttt{Apache commons-math}.}
    \label{fig:mutant:example}
\end{figure*}

As an example of a mutant killed only by $T_{\cling}$, Figure \ref{fig:mutant:example:callee} illustrates one of the mutants in method  \texttt{evaluateStep} in class \texttt{Switch\-Sta\-te} (callee class) from the \textit{Apache commons-math} project. 
This mutant is produced by an integration-level mutation operator (\textit{RetStaRep}) that replaces a boolean return value by \texttt{true}.
Method \textit{evaluateStep} is called from the method \texttt{evaluateStepC} (Figure~\ref{fig:mutant:example:caller}) declared in \texttt{SwitchingFunctionsHandler} (caller class). Method \texttt{evaluateStepC} must return false if it calls the callee class in a certain situation: (i) the variable \texttt{first} in the caller class is null, and (ii) the callee method returns false because of the execution of line 12 in Figure~\ref{fig:mutant:example:callee}. 

The unit test suites generated by \evosuite targeting \texttt{Switch\-Sta\-te} ($T_{EvoE}$) or class \texttt{Swi\-tching\-Fun\-ctions\-Hand\-ler} ($T_{EvoR}$) both cover the mutant but do not kill it. 
$T_{EvoE}$ easily cover the mutant statement, but it does not have any assertion to check the return value.
$T_{EvoR}$ also covers this statement by calling the right method in \texttt{SwitchingFunctionsHandler}. However, as is depicted by Figure~\ref{fig:mutant:example}, both methods in caller and callee class have multiple branches. So, $T_{EvoR}$ covers the mutant from another path, which does not reveal the change in the boolean return value.

\begin{lstlisting}[frame=tb,
    caption={\cling test case killing mutant in Figure~\ref{fig:mutant:example}.},
    label=list:clingTest,
    captionpos=t,
    numbers=left,
    float=t,
    belowskip=-2.5em,
    firstnumber=1]
    public void test07()  throws Throwable  {
        [...]
        boolean boolean1 = switchingFunctionsHandler0.evaluateStepC(stepInterpolator0);
        assertTrue(boolean1 == boolean0);
        assertFalse(boolean1);
    }
    
\end{lstlisting}

In contrast, this mutant is killed by $T_{\cling}$, targeting \texttt{SwitchingFunctionsHandler} and \texttt{SwitchState} as the caller and callee classes, respectively (Listing \ref{list:clingTest}).
According to the assertion in line 5 of this test case, \texttt{swit\-ching\-Func\-tions\-Hand\-ler0.e\-va\-lu\-ate\-Step} must return false. However, the mutant changes the returned value in line 7 of the caller class (Figure \ref{fig:mutant:example:caller}), and thereby the true branch of the condition in line 7 is executed. This true branch changes the value of variable \texttt{first} from null to a non-null value. Hence, the \texttt{evaluateStep} method in the caller class returns true in line 12. So, the assertion in the last line of the method in Listing~\ref{list:clingTest} kills this mutant.

\smallskip\noindent\fbox{\parbox{0.475\textwidth}{
\textbf{Summary (RQ2.2).}  
The test suite generated by \cling for a caller $R$ and callee $E$, can kill \textit{different} mutants than unit and random test suites for $E$, $R$ or their union, increasing the mutation score on average by 13.0\%, 10.4\%, and 7.7\%, respectively, for \evosuite, and 13\% for \randoop, with outliers well above 50\%. 
Our analysis indicates that many of the most frequently killed mutants are produced by integration-level mutation operators.
}}

\subsection{Integration Faults Exposed by \integration (RQ3)}
\label{sec:results:rq3}

In our experiments, \integration generates 50 test cases that triggered unexpected exceptions in the subject systems. 
None of those  exceptions were observed during the execution of the test cases generated by \evosuite and \randoop.

\begin{table}[t]
	\center
    \caption{Categorization and number (\textbf{\#}) of the fault revealing test cases.} 
    \label{tab:testcategory}
    \begin{footnotesize}
        \begin{tabular}{p{12mm} r p{55mm}}
            \toprule
            \textbf{Category} & \textbf{\#} & \textbf{Description} \\
            \midrule 
            \textbf{Confirmed}  & 7 & The test case exposes a fault that has been fixed (\eg by updating the code or the documentation), or has been marked as such in the source code (\eg using a comment).\\
            \textbf{Pending} & 4 & The test case potentially exposes a fault that has not been fixed.\\
            \textbf{Deprecated} & 14 & The test case is not relevant anymore as the source code it executes has been deleted from the project (\eg in the case of a deprecated method).\\
            \bottomrule 
        \end{tabular}
    \end{footnotesize}
\end{table} 

\newsavebox\exceptionbox
\begin{lrbox}{\exceptionbox}
\begin{minipage}{0.45\textwidth}
\begin{lstlisting}[
    numbers=left,
    firstnumber=1]
java.lang.NullPointerException:
 at [..].JSType.isEmptyType([..]:159)
 at [..].JSType.testForEqualityHelper([..]:666)
 at [..].JSType.testForEquality([..]:655)
 at [..].NumberType.testForEquality([..]:63)
 at [..].JSType.getTypesUnderInequality([..]:962)
 at [..].UnionType.getTypesUnderInequality([..]:486)
 at [..].JSType.getTypesUnderInequality([..]:957)
 at [..].UnionType.getTypesUnderInequality([..]:486)
\end{lstlisting}
\end{minipage}
\end{lrbox}

\newsavebox\testcasebox
\begin{lrbox}{\testcasebox}
\begin{minipage}{0.45\textwidth}
\begin{lstlisting}[
    numbers=left,
    firstnumber=1]
public void testFraction() {
    [...]
    UnionType unionType0 = new UnionType((JSTypeRegistry) null, immutableList0);

    // Undeclared exception!
    unionType0.getTypesUnderInequality(unionType0);
}
\end{lstlisting}
\end{minipage}
\end{lrbox}

\begin{figure*}[t]
    \centering
    \subfloat[\cling test case triggering the crash in Figure~\ref{list:divzero}.]{\usebox\testcasebox\label{list:testdivzero}}
    \hfil
    \subfloat[Exception captured only by \integration.]{\usebox\exceptionbox\label{list:divzero}}
    \caption{Example of test case generated by \cling and exposing a fault in the \textit{Closure} project.}
    \label{fig:divzeroexample}
\end{figure*}

The first and second author independently performed a manual root cause analysis for all 50 unexpected exceptions to check if they actually stemmed from an integration-level fault. For this analysis, we check the API documentation to see if the generated test cases break any precondition. We indicated a test case as a fault revealing test if it does not violate any precondition according to the documentation, and it truly exposes an issue about the interaction between the caller and callee class. We found that out of the 50 test cases generated by \cling, 25 are fault revealing. The remaining 25 test cases trigger exceptions expected according to the documentation (5 tests), violate preconditions specified in the documentation (14 tests) or in the existing tests (1 test), return a wrong value for a method call on a mocked object (3 tests), or do not actually expose an issue between the caller and the callee class (2 tests).

To analyze if developers have already identified the faults in the following commits, the first three authors analyzed the code history of the classes involved in the detected faults. In this analysis, we manually checked all of the modifications made to the involved classes to see if the faults are fixed. Based on this analysis, we classify the 25 fault revealing test in one of the categories reported in Table~\ref{tab:testcategory}.
According to this Table, seven faults (found only by tests generated via \cling) were detected, confirmed, and fixed by developers in the next commits. We describe hereafter a representative example of these faults. 
The detailed descriptions of the analysis for all 25 fault revealing test  cases are available in our replication package \cite{Derakhshanfar2020}.\footnote{Also available online at \url{https://github.com/STAMP-project/Cling-application/blob/master/data_analysis/manual-analysis/failure-explanation.md}.}

\smallskip \textbf{Example.} 
To illustrate the type of problem detected by \integration, consider the generated test case in Figure~\ref{list:testdivzero} and the induced stack trace (for a \texttt{NullPointerException}) in Figure~\ref{list:divzero}.\footnote{The details are available at \url{https://github.com/STAMP-project/Cling-application/blob/master/data_analysis/manual-analysis/failure-explanation.md\#st28}} 
This test is produced by \cling for classes \texttt{UnionType} (caller class) and \texttt{JSType} (callee class). In this scenario, the \texttt{UnionType} is a sub-class of \texttt{JSType}.
The test (Figure~\ref{list:testdivzero} Line 3) instantiates a \texttt{UnionType} object and passes a \texttt{null} value for the first parameter of its constructor. This constructor sets the value of a local variable (\texttt{registry}) to the value passed as the first parameter of the constructor (here, \texttt{null}).
After instantiating \texttt{UnionType}, the generated test calls \texttt{getTypesUnderInequality} (Figure~\ref{list:testdivzero} Line 6), which in turns indirectly calls \texttt{isEmptyType} in the superclass. The \texttt{isEmptyType} method tries to use the attribute \texttt{registry}. Since this attribute is \texttt{null}, calling \texttt{getTypesUnderInequality} leads to a \texttt{NullPointerException}. No indication in the documentation specifies that the \texttt{registry} parameter should not be \texttt{null}, and no checks are done on the value of the input parameters.

By reviewing the code history of \texttt{UnionType} class, we observed that this fault has been fixed.\footnote{The fixing commit is \url{https://github.com/google/closure-compiler/commit/cfc0fab3dc2be49692a4fe9162b8095c934f6c41}.} 
A \texttt{UnionType} should be instantiated only by a \texttt{UnionTypeBuilder} to ensure that it is instantiated properly, but this was not enforced in the source code nor documented in the class. 
The fixing commit message indicates that it refactors the \say{\texttt{Union\-Type\-Buil\-der} into \texttt{Union\-Type.Buil\-der}, a nested class of \texttt{Union\-Type}} to \say{better reflect the entangled nature of the two classes.}
Concretely, the commit 
(i) refactors the \texttt{UnionTypeBuilder} class into \texttt{UnionType.Builder}, a nested class of \texttt{UnionType}; 
(ii) makes \texttt{UnionType}'s constructor private; 
and (iii) updates the \texttt{UnionType} constructor's documentation to indicate that this class has to be instantiated using its builder.

We also analyzed the tests generated by baseline tools for this case to understand why this specific fault is substantially less likely to be captured by \evosuite and \randoop. For \evosuite, since the tool concentrates on coverage of a single class, the tests generated by \evosuite only concentrate on covering the branches in the given class under test. So, in the \evosuite test generation process, tests cases that achieve higher branch coverage have higher priority than this failure capturing test case. This prioritization leads \evosuite to exclude this test case from the most optimized solutions during the search process. 
In contrast with \evosuite, \cling's search objectives (\ie CBC coverage) are designed to exercise the interactions between given class pairs and thereby give a higher priority to failures that can be captured in this interaction.
Moreover, since \randoop gets a set of classes under test (\ie classes that are direct or indirect dependencies of the given caller and callee classes), it has a higher search space to explore.
In this case, \randoop generates tests using 30 classes indicated by jdeps (25 classes from the project under test and five from Java). In total, these classes contain 867 visible (non-private) methods.
Also, the tests generated by \randoop initialize and use many objects, and hence the length of test cases are relatively higher than test cases generated by \evosuite and \cling. Consequently, by looking at tests generated by \randoop, we can see that this tool generates many test cases that lead to higher coverage in the given set of classes but could not explore the particular part of the search space to capture this fault in the given time budget. However, theoretically, by giving enough time budget to \randoop, this tool should be able to cover this failure. In contrast, since \cling focuses on the interactions between two given classes, thereby having a smaller search space, it manages to capture this failure in a shorter time (\ie 5 minutes).

\smallskip\noindent\fbox{\parbox{0.475\textwidth}{
\textbf{Summary (RQ3).}  
Our manual analysis indicates that \integration-based automated testing of $\langle$caller, callee$\rangle$ class pairs can expose actual problems that are not found by unit testing either the caller or callee class individually. These problems relate to conflicting assumptions on the safe use of methods across classes (\eg due to undocumented exception throws, implicit assumptions on parameter values, \etc). Several of these faults are identified, confirmed, and fixed later by developers in subsequent commits.
}}

\section{Discussion}
\label{sec:discussion}

\subsection{Applicability}

The CBC criterion and its implementation in \cling consider pairs of classes and targets the integration between them. We did not propose any procedure for selecting pairs of classes to give in input to \integration. 
Since the technique requires pairs of classes to test, it would be time-consuming and tedious for developers to manually collect and provide the class pairs. Hence, we suggest using an automated process for class pair selection, as well. In this study, we implemented a tool that automatically analyzes each class pair to find the ones with high cyclomatic complexity and coupled branches (according to the CBC criterion defined in this article). This procedure is explained in Section \ref{sec:setup:subselection}. 

Besides, our approach can be further extended by incorporating automated integration test prioritization approaches and selecting classes to integrate according to a predefined ordering \cite{Wang2010, Hashim2005, Vergilio2012, Bansal2009, JIiang2019, Mariani2016, Guizzo2015, Abdurazik2009, DaVeigaCabral2010, Briand2003a}.
So, the end-to-end process of generating test for class integrations can be automated to require a minimal manual effort from the developer.

Moreover, since it is easier for developers to handle and integrate generated test cases in continuous integration, the number of tests generated by test generation approaches is also playing a crucial role in the effectiveness and applicability of techniques. Although \cling generates test cases that kill mutants and capture integration-level failures that cannot be covered by unit and random testing, this approach generates less test cases compared to \evosuite and \randoop. In total, \cling generated 64,537 test cases in this experiment. This number is lower than \evosuite (183,795 test cases) and \randoop (29,536,928 test cases). 


\subsection{Test generation cost}

One of the challenges in automated class integration testing is detecting the integration points between classes in a SUT. The number of code elements (\eg branches) that are related to the integration points increases with the complexity of the involved classes. Finding and testing a high number of integration code targets increases the time budget that we need for generating integration-level tests. 

With CBC, the number of coupled branches to exercise is upper bounded to the cartesian product between the branches in the caller $R$ and the callee $E$. Let $B_R$ be the set of branches in $R$ and $B_E$ the set of branches in $E$, the maximum number of coupled branches $CB_{R,E}$ is $B_R \times B_E$. In practice, the size of $CB_{R,E}$ is much smaller than the upper bound as the target branches in the caller and callee are subsets of $R$ and $E$, respectively. Besides, CBC is defined for pairs of classes and not for multiple classes together. This substantially reduces the number of targets we would incur when considering more than two classes at the same time.

While a fair amount of the test generation process can be automated, multiple instances of this approach can be executed simultaneously, and thereby, this approach can be used to generate test suites for a complete project at once in a reasonable amount of time. For instance, in this study, we managed to test each of the 140 class pairs with \cling for 20 times in less than a day thanks to a parallelization of the executions.

Finally, we have used a five minutes time budget to test each class pair's interactions. Since \cling considers each coupled branch as an objective for the search process, we could have defined a different search budget per pair, depending on the number of objectives. Similarly to \evosuite and \randoop, the outcome of \cling may differ depending on the given time budget. Defining the best trade-off between the search-budget and effectiveness of the tests generated using \cling is part of our future work. 

\subsection{Effectiveness}

To answer \textbf{RQ2}, we analyzed the set of mutants that are killed by \cling (integration tests), but not by the unit and random test suites for the caller and callee separately (boxes labeled with $T_{EvoE+R}$ and $T_{Ran}$ in Figure~\ref{fig:mutation:diff:boxplot}). The test suite $T_{\cling}$ was generated using a search budget of five minutes. Similarly, the unit-level suites were generated with a search budget of five minutes for each caller and callee class separately. Therefore, the total search budget for unit test generation ($T_{EvoE+R}$) is twice as large: 10 minutes, which corresponds to the time budget allocated to \randoop ($T_{Ran}$) as it generates random tests for both the caller and the callee. Despite the larger search budget spent on unit and random testing, there are still mutants and faults detected only by \cling and in less time.
It is worth mentioning that, theoretically, all of these approaches might capture these failures with an infinite time budget. The point is that Cling can capture these failures faster, thanks to the CBC criterion.

\textbf{The CBC criterion and its implementation in \cling are not an alternative to unit or random testing}. In fact, integration test suites do not subsume unit-level and random suites as the different types of suites focus on different aspects of the system under test. Our results (\textbf{RQ2}) confirm that integration and unit and random testing are complementary. Indeed, some mutants can be killed exclusively by unit or random test suites: \eg the overall mutation scores for the unit tests $T_{EvoE}$ and $T_{EvoR}$, and random tests $T_{Ran}$ are larger than the overall mutation scores of \cling. This higher mutation score is expected due to the larger branch coverage achieved by the unit and random tests (\ie coverage is a necessity but not a sufficient condition to kill mutant). 

Instead, the CBC criterion and its implementation in \cling focuses on a subset of the branches in the units (caller and callee), but target the integration between them more extensively. In other words, the search is less broad (fewer branches), but more in-depth (the same branches are covered multiple times within different pairs of coupled branches). This more in-depth search allows killing mutants that could not be detected by satisfying unit-level criteria.

Furthermore, our results in \textbf{RQ3} indicate that CBC and its implementation in \cling steer us toward finding bugs that are not detectable by other tests. In Section~\ref{sec:results:rq3}, we have shown that the tests generated by \cling capture exceptions, which are not detectable by unit or random tests. We have carefully performed an extensive manual analysis on these stack traces to identify whether they expose software faults. According to this manual analysis, we have detected 25 failures. Finally, for external confirmation, we have investigated if these 25 faults are identified and fixed by developers in the subsequent commits. The results of our investigation have confirmed that developers have actually fixed some of the faults in the following commits. To demonstrate the impact of the \cling in finding bugs, we have presented an example in Section~\ref{sec:results:rq3}. Moreover, the other faults, which were confirmed by our investigation, are available in our replication package~\cite{Derakhshanfar2020}. While our evaluation pointed to 25 real faults, we have not yet applied \cling in a live setting in a currently active project. Doing so requires a project that does intensive (unit) testing already, and whose developers are interested in exploring issues raised by tests dedicated to exercising various inter-class interactions. As part of our future work, we intend to set up and conduct such a (longitudinal) study.

\section{Threats to validity}
\label{sec:threats}

\smallskip \textbf{Internal validity.}
Our implementation  may contain bugs. We mitigated this threat by reusing standard algorithms implemented in \evosuite, a widely used state-of-the-art unit test generation tool. And by unit testing the different extensions (described in Section \ref{sec:implementation}) we have developed. 

To take the randomness of the search process into account, we followed the guidelines of the related literature \cite{Arcuri2014} and executed \cling, \evosuite, and \randoop \nrun times to generate the different test suites ($T_{\cling}$, $T_{EvoE}$, $T_{EvoR}$, $T_{RanE}$, and $T_{RanR}$) for the 140 caller-callee classes pairs. 
We have described how we parametrize \cling, \evosuite, and \randoop in Sections \ref{sec:cling} and \ref{sec:evaluation}. We left all other parameters to their default value, as suggested by related literature \cite{Arcuri2013, Panichella2015, Shamshiri2018}.

\smallskip \textbf{External validity.} 
We acknowledge that we report our results for only five open-source projects. However, we recall here their diversity and broad adoption by the software engineering research community. 
We also did not use the latest version of those five projects. On the one hand, it prevented us from reporting potential faults to the developers, which could have provided anecdotal evidence of the capability of the approach to find faults, but would have not provided any information in case of a rejection of a pull request by the developers. On the other hand, it allowed us to investigate the history of the code base and identify whether developers fix these faults, which were identified in our study, in the further commits. Additionally, considering the broad adoption of the projects by the software engineering community, it enables comparisons with the state-of-the-art and future approaches. 

\smallskip \textbf{Construct validity.} 
The identification and analysis of the integration faults done in \textbf{RQ3} have been performed by the first and second authors independently. The subsequent code history analysis and categorization have been performed by the three first authors independently. Each documented analysis was reviewed by one of the other authors involved.

\smallskip \textbf{Reproducibility.} 
We provide \cling as an open-source publicly available tool as the data and the processing scrips used to present the results of this paper.\footnote{\url{https://github.com/STAMP-project/Cling-application}} Including the subjects of our evaluation (inputs) and the produced test cases (outputs). The full replication package has been uploaded on Zenodo for long-term storage \cite{Derakhshanfar2020}.

\section{Conclusion and future work}
\label{sec:future-conclusion}

In this paper we have introduced a testing criterion for integration testing, called the \textit{Coupled Branches Coverage} (CBC) criterion. 
Unlike previous work on class integration testing focusing on (costly) data-flow analysis, CBC relies on a (lighter) control flow analysis to identify couples of branches between a caller and a callee class that are not trivially executed together, resulting in a lower number of test objectives. 

Previous studies have introduced many automated unit and system-level testing approaches for helping developers to test their software projects. However, there is no approach to automate the process of testing the integration between classes, even though this type of testing is one of the fundamental and labor-intensive tasks in testing. To automate the generation of test cases satisfying the CBC criterion, we defined an evolutionary-based class integration testing approach called \integration.

In our investigation of 140 branch pairs, collected from 5 open source Java projects, we found that \integration has reached an average CBC score of 49.1\% across all classes, while for some classes we have reached 90\% coverage. More tangibly, if we consider mutation coverage and compare automatically generated random and unit tests with automatically generated integration tests using the \integration approach, we find that our approach allows to kill 7.7\% (resp. 13\%) of mutants per class that cannot be killed by tests generated with \evosuite (resp. \randoop). Finally, we identified 25 faults causing system crashes that could be evidenced only by the generated class-integration tests. 

The results indicate a clear potential application perspective, more so because our approach can be incorporated into any integration testing practice. Additionally, \integration can be applied in conjunction with other automated unit and system-level test generation approaches in a complementary way.

From a research perspective, our study shows that \integration is not an alternative for unit or random testing. However, it can be used for complementing unit testing for reaching higher mutation coverage and capturing additional crashes which materialize during the integration of classes. These improvements of \integration are achieved by the key idea of using existing usages of classes in calling classes in the test generation process. 

For now, \integration only tests the call-coupling between classes. In our future work, we will extend our approach to explore how other types of coupling between classes (\eg parameter coupling, shared data coupling, and external device coupling) can be used to refine the couples of branches to target. 
Indeed, our study indicates that despite the effectiveness of \integration in complementing unit tests, lots of objectives (coupled branches) remain uncovered during our search process. Hence, in future studies, we will enhance the detection of infeasible branches to remove them from the search objectives and perform a fitness landscape analysis of the search process to identify potential bottlenecks. 

Finally, this study mostly focuses on examining the results of this approach on coupled branches coverage, mutation coverage, and detected faults. In our future work, we will explore how \cling can be effectively integrated with a development lifecycle (for instance, in a continuous integration process) and how automatically generated class integration tests can help developers to detect potential faults and debug their software.

In this study, we have evaluated \cling against state-of-the-art test generation tools (\ie \evosuite and \randoop). In our future work, we would like to compare the tests generated by \cling with the manually written tests. Also, since CBC is a new criterion, we aim to perform another study to investigate how well the class integration tests written by developers cover CBC targets.

Finally, since this paper is the first step toward generating class integration tests, we only collected the call-sites from the static analysis. However, a dynamic analyzer is able to detect more call-sites, and thereby CLING can generate more tests that cover class interactions that can only be identified dynamically.

\ifCLASSOPTIONcompsoc
  \section*{Acknowledgments}
\else
  \section*{Acknowledgment}
\fi

This research was partially funded by the EU Project STAMP ICT-16-10 No.731529, the EU Horizon 2020 H2020-ICT-2020-1-RIA ``COSMOS'' project (No.957254), and the NWO Vici project ``TestShift'' (No. VI.C.182.032).




%
\balance
\bibliographystyle{IEEEtran}
\bibliography{IEEEabrv,bibliography}

\end{document}